\documentclass[preprintnumbers,amsmath,amssymb,aps]{revtex4}
\usepackage{graphicx}
\usepackage{latexsym}
\usepackage[dvips]{color}

\def\Tr{{\rm Tr}}

\begin{document}
\title{Renormalization of Bilinear Quark Operators for Overlap
Fermions}
\author{Thomas DeGrand}
\author{Zhaofeng Liu}
\affiliation{
Department of Physics, University of Colorado,
        Boulder, CO 80309 USA}
\begin{abstract}
We present non-perturbative renormalization constants of fermionic
bilinears on the lattice in the quenched approximation at $\beta=6.1$
using an overlap~\cite{neub} fermion action with hypercubic(HYP)-blocked links. 
We consider the effects of the exact zero
modes of the Dirac operator and find they are important in calculating
the renormalization constants of the scalar and pseudoscalar density.
The results are given in the RI' and $\overline{MS}$ schemes and compared
to the perturbative calculations.
\end{abstract}
\maketitle
\section{Introduction}\label{Intr}
This paper describes the computation of matching factors for
converting lattice calculations of matrix elements of currents to the
corresponding values measured in a continuum
 $\overline{MS}$ scheme. The lattice action uses
 overlap fermions~\cite{neub} and HYP-blocked links~\cite{hase}.

We use the nonperturbative methodology 
introduced in Ref.~\cite{npr1}. The proposed renormalization scheme
is one which can be implemented not only in lattice Monte Carlo
simulation but also in continuum perturbation theory. Thus, the
conversion of lattice results to a more conventional scheme such as
$\overline{MS}$ is possible. In this scheme, the matrix element of a
bilinear quark operator $O_{\Gamma}=\bar{\psi}\Gamma\psi$ between
quark fields at certain momentum $p^2=\mu^2$ is computed and matched
to the corresponding tree level matrix element. i.e. the renormalization
condition is
\begin{equation}
Z_\Gamma\langle p|O_\Gamma|p\rangle|_{p^2=\mu^2}=\langle
p|O_\Gamma|p\rangle_{tree}.
\label{recond}
\end{equation}
Here $\Gamma$ can be any combination of Dirac matrices.
This method is supposed to work when $\mu$ satisfies
\begin{equation}
\Lambda_{QCD}\ll\mu\ll1/a.
\end{equation}
The discretization effects are under control if the renormalization
scale $\mu$ is much smaller than the lattice cut off $1/a$.
$\Lambda_{QCD}\ll\mu$ guarantees that the non-perturbative effects are
ignorable.

There have been many calculations using this scheme.
Ref.~\cite{npr1} used improved Wilson fermions, Ref.~\cite{gime}
used both the Wilson and the tree level improved SW-Clover fermion
action in the quenched approximation, Ref.~\cite{gock} worked with standard
Wilson fermions($r=1$) in the quenched approximation, Ref.~\cite{gatt} used
chirally improved lattice fermions in the quenched approximation.
Here,
we  use overlap fermions~\cite{neub} in the quenched 
approximation. Specifically, we work with overlap fermions
built from a "kernel action" with nearest and next-nearest neighbor
fermionic interactions~\cite{ovaction} and 
hypercubic(HYP)-blocked links~\cite{hase}.
Overlap fermions respect chiral symmetry on the lattice
via the Ginsparg-Wilson relation~\cite{gw}, while for Wilson-type
fermions, the Wilson term breaks the
chiral symmetry explicitly. Chirally improved fermions only
obey the Ginsparg-Wilson relation approximately.

Lattice perturbation theory is probably the most often
used way  to calculate the renormalization
factors. However, the convergence of the perturbative series is often
not satisfying. To improve the convergence of the series, Lepage and
Mackenzie proposed a tadpole improved perturbation theory~\cite{tadpole}.
Nevertheless, lattice perturbation series rarely extend beyond the
one-loop level, which is an important source of uncertainty in the
extraction of physical results.

One-loop perturbative calculations of the matching coefficients between
matrix elements measured in lattice simulations and their equivalent
$\overline{MS}$ values for the same overlap fermions and HYP-blocked links
that we use here were done in Ref.~\cite{degr}. Those perturbative results
turned out to be quite close to unity, and they were used in computing
the Kaon B parameter~\cite{degrb}. This work will give a non-perturbative
check of the matching coefficients.

Perturbative results of the matching coefficients for other actions
also using HYP-blocked links or similar gauge connections
were presented in \cite{lees}. They show the same behavior that the
matching coefficients are quite close to unity. Our results may be useful to
others doing simulations with HYP links, to give an idea how trustworthy
perturbation theory is.

In Eq.~(\ref{recond}), $\Gamma$ can be any combination of Dirac
matrices. We will consider the cases $\Gamma=I$,
$\gamma_5$, $\gamma_\mu$ and $\gamma_\mu\gamma_5$,
which we will denote S, P, V and A respectively.
Chiral symmetry implies several
relations between renormalization constants for overlap fermions, 
in particular $Z_S=Z_P$ and $Z_V=Z_A$.

The paper is organized as follows: In Section \ref{Meth}, 
we briefly discuss
the non-perturbative renormalization method~\cite{npr1}, the overlap
action and how we deal with the zero modes of the Dirac operator.
Numerical results are given in Section \ref{NumRes}. 
The formulas for conversion to
$\overline{MS}$ scheme are recapitulated in Section \ref{Conv}. We will 
compare our results with perturbative calculations in Section \ref{Comp}
and conclude in Section \ref{Summ}.

\section{Methodology}\label{Meth}
The following is a brief summary of the method from
Ref.~\cite{npr1,gatt}, a short description of the overlap action
we used (for a detail description, see Ref.~\cite{ovaction}) and
how we deal with the zero modes of the Dirac
operator.
For convenience, the lattice spacing $a$ is set to be
one.

From Eq.~(\ref{recond}), we have
\begin{equation}
Z_\Gamma\frac{1}{12}\Tr\left.\left[\langle p|O_\Gamma|p\rangle\langle p|
O_\Gamma|p\rangle_{tree}^{-1}\right]\right|_{p^2=\mu^2}=1.
\label{rencon}
\end{equation}
Here $\frac{1}{12}$ comes from the fact that the trace is over color
and spin space.
Since
\begin{equation}
\langle p|O_\Gamma|p\rangle=Z_q\Lambda_\Gamma(p),
\label{matrixamp}
\end{equation}
we obtain
\begin{equation}
Z_\Gamma=\left.\frac{12}{Z_q\Tr\left[\Lambda_\Gamma(p)
\langle p|O_\Gamma|p\rangle_{tree}^{-1}\right]}\right|_{p^2=\mu^2}.
\label{zgamma}
\end{equation}
Here $Z_q$ is the quark field renormalization constant 
(The bare field $\psi_0=Z_q^{1/2}\psi$) and
$\Lambda_\Gamma(p)$ is the amputated Green function
\begin{equation}
\Lambda_\Gamma(p)=S^{-1}(p)G_\Gamma(p)S^{-1}(p),
\label{ampgreen}
\end{equation}
where $S(p)$ is the quark propagator.
Eq.~(\ref{zgamma}) is the formula we will use to calculate $Z_\Gamma$.

$Z_q$ is obtained by comparing the quark propagator to the free
lattice propagator (the RI' scheme):
\begin{equation}
Z_q^{RI'}=\left.\frac{1}{12}\Tr[S(p)D_f^{ov}(p)]\right|_{p^2=\mu^2},
\label{zqfactor}
\end{equation}
where $D_f^{ov}(p)$ is the free lattice overlap Dirac operator. 
(Our $Z_q$ is the inverse of the quark field renormalization constant
in Ref.~\cite{gatt}.)

The Green function $G_\Gamma(p)$ is determined in the following way.
\begin{eqnarray}
G_\Gamma(p)&=&\sum_{x,y}e^{-ip\cdot
(x-y)}\langle\psi(x)O_\Gamma(0)\bar{\psi(y)}\rangle \nonumber\\
&=&\sum_{x,y}e^{-ip\cdot (x-y)}\frac{1}{N}\sum_{i=1}^N S_i(x|0)\Gamma
S_i(0|y) \nonumber\\
&=&\frac{1}{N}\sum_{i=1}^N\left(\sum_x S_i(x|0)e^{-ip\cdot x}\right)
\Gamma\left(\sum_y S_i(0|y)e^{ip\cdot y}\right),
\end{eqnarray}
where $N$ is the number of gauge configurations. Using
$S_i(x|y)=\gamma_5S_i(y|x)^\dagger\gamma_5$ and 
$S_i(p|0)=\sum_xS_i(x|0)e^{-ip\cdot x}$, we have
\begin{equation}
G_\Gamma(p)=\frac{1}{N}\sum_{i=1}^NS_i(p|0)\Gamma\gamma_5S_i^\dagger(p|0)
\gamma_5.
\label{green}
\end{equation}
The quark propagator in momentum space is given by
\begin{equation}
S(p)=\frac{1}{N}\sum_{i=1}^NS_i(p|0).
\label{sp}
\end{equation}
$S_i(x|0)$ is computed on the lattice with a point source
\begin{equation}
\sum_xD^{ov}(z,x)S_i(x|0)=\delta_{z,0}.
\end{equation}
(In Ref.~\cite{gock} and Ref.~\cite{gatt}, momentum sources were used.)
At tree level, $\langle p|O_\Gamma|p\rangle_{tree}=\Gamma$. Therefore, every
quantity on the right hand side of Eq.~(\ref{zgamma}) is known and then we
can obtain $Z_\Gamma$. For $\Gamma=\gamma_\mu,\gamma_\mu\gamma_5$,
the index $\mu$ is averaged under the trace in Eq.~(\ref{zgamma}).

We fix the gauge to Landau gauge.
Uncertainty due to Gribov copies is not investigated here. It has been
discussed in Ref.~\cite{gatt,paci,giusti,giust}. The effect was found
to be negligible in current lattice simulations.

The overlap
action that we use is described with detail in Ref.~\cite{ovaction}, 
which uses a "kernel"
action with nearest and next-nearest neighbor couplings. 
The massless overlap Dirac operator is
\begin{equation}
D(0)=x_0\left(1+\frac{z}{\sqrt{z^\dagger z}}\right),
\end{equation}
where $z=d(-x_0)/x_0=(d-x_0)/x_0$ and $d(m)=d+m$ is the massive
Dirac operator for mass $m$.
The overall multiplicative factor of $x_0$
is a useful convention so that when $D(0)$ is expanded for small
$d$, $D\approx d$.

The massive overlap Dirac operator is defined as
\begin{equation}
D(m)=\left(1-\frac{m}{2x_0}\right)D(0)+m.
\end{equation}
In a background gauge field carrying a topological charge $Q$,
$D(0)$ will have $|Q|$ pairs of real eigenmodes with eigenvalues
$0$ and $2x_0$. In computing propagators, it is convenient to
clip out the eigenmode with real eigenvalue $2x_0$, and to define
the subtracted propagator as
\begin{equation}
\tilde{D}(m)^{-1}=\frac{1}{1-\frac{m}{2x_0}}\left[
D(m)^{-1}-\frac{1}{2x_0}\right].
\end{equation}
This also converts local currents into order $a^2$ improved 
operators~\cite{Capitani:1999uz}.
Then the free lattice overlap Dirac operator $D^{ov}_f(p)$
used in Eq.~(\ref{zqfactor}) is just $\tilde{D}(m)$ in the
momentum space.

The HYP-blocked links are constructed in three steps~\cite{hase}.
The parameters $\alpha_1$, $\alpha_2$ and $\alpha_3$ in our simulation
have the favored values of 
Ref.~\cite{hase}: 0.75, 0.6 and 0.3 respectively.

A finite volume artifact we encounter in this quenched simulation is 
the presence of exact zero modes of the Dirac operator. The zero mode
contribution (with positive chirality) in the propagator
$S_i(p|0)$ on a configuration with $Q\neq0$ takes the form
\begin{equation}
\frac{1}{m}\left(\begin{array}{cc}
|\phi_0(p)\rangle\langle\phi_0(p)|&0\\0&0
\end{array}\right)\equiv\frac{1}{m}S_0
\label{0mode}
\end{equation}
in $\gamma_5$ diagonal basis. Here $|\phi_0(p)\rangle$ is the Fourier
transform of the zero mode wave function $|\phi_0(x)\rangle$. 
Since the zero modes are localized
in space, $|\phi_0(p)\rangle$ will peak at low $p$. 
These zero modes do not resemble free field modes. Implicit in the
RI' scheme analysis is the idea that at big $\mu$, lattice propagators
resemble continuum ones. Zero modes clearly do not. In Section
\ref{NumRes}, we will find zero modes make a large contribution to
$Z_S$ and $Z_P$. We believe this is because our lattice is not large.

The following little parametrization illustrates our expectations
of the effects of zero modes:
$S_i(p|0)$ is the sum of the zero mode
contribution and the non-zero mode contribution $S_n$,
\begin{equation}
S_i(p|0)=\frac{1}{m}S_0+S_n.
\end{equation}
Therefore in Eq.~(\ref{green})
\begin{equation}
S_i(p|0)\Gamma\gamma_5S_i^{\dagger}(p|0)\gamma_5=
\frac{1}{m^2}S_0\Gamma\gamma_5S_0^\dagger\gamma_5+\frac{1}{m}
(S_0\Gamma\gamma_5S_n^\dagger\gamma_5+S_n\Gamma\gamma_5S_0^\dagger\gamma_5)
+S_n\Gamma\gamma_5S_n^\dagger\gamma_5,
\end{equation}
and then $G_\Gamma(p)$ can be written in the form
\begin{equation}
G_\Gamma(p)=\frac{1}{m^2}G_2+\frac{1}{m}G_1+G_0,
\end{equation}
where the subscript counts the number of zero modes:
 $G_0$ contains no zero mode contribution.
The quark propagator averaged over all configurations and
its inverse, if expanded for small $m$, are
\begin{eqnarray}
S(p)&=&\frac{1}{m}\bar{S}_0+\bar{S}_n,\nonumber\\
S^{-1}(p)&=&m\bar{S}_0^{-1}-m^2\bar{S}_0^{-1}\bar{S}_n\bar{S}_0^{-1}+\cdots.
\end{eqnarray}
Thus the amputated Green function
\begin{eqnarray}
\Lambda_\Gamma(p)&=&S^{-1}(p)G_\Gamma(p)S^{-1}(p)\nonumber\\
&=&\left(\frac{1}{m}\bar{S}_0+\bar{S}_n\right)^{-1}
\left(\frac{1}{m^2}G_2+\frac{1}{m}G_1+G_0\right)
\left(\frac{1}{m}\bar{S}_0+\bar{S}_n\right)^{-1}\nonumber\\
&\rightarrow&\bar{S}_0^{-1}G_2\bar{S}_0^{-1}\quad\mbox{when $m$ is
small}.
\end{eqnarray}
So, if the zero modes affect our calculation of $Z_\Gamma$ 
(Eq.~(\ref{zgamma})), the effect should be evident at small momentum and small
quark mass. Unfortunately, for us, $\mu=2$~GeV, where we will match, is
rather small momentum.

We examine
two solutions to those zero modes. One solution is to explicitly 
subtract the contribution of the zero modes in the quark propagator.
We would prefer not to do this.
The other solution is to use combination of scalar and pseudoscalar
densities or combination of vector and axial vector currents so that
the zero mode contributions are suppressed in $Z_\Gamma$. We can
do the latter because the overlap fermion respects chiral symmetry on
the lattice.

For scalar and pseudoscalar, we can use Eq.~(\ref{matrixamp}) to 
rewrite Eq.~(\ref{recond}) as
\begin{equation}
Z_SZ_q\Lambda_I(p^2=\mu^2)=I
\end{equation}
and
\begin{equation}
Z_PZ_q\Lambda_{\gamma_5}(p^2=\mu^2)=\gamma_5.
\end{equation}
If $Z_{SP}\equiv Z_S=Z_P$, then we have
\begin{equation}
Z_{SP}Z_q(\Lambda_I\pm\Lambda_{\gamma_5})=I\pm\gamma_5,
\end{equation}
and thus
\begin{equation}
Z_{SP}Z_q\frac{1}{12}\Tr(\Lambda_I\pm\Lambda_{\gamma_5})=1.
\label{zspeq1}
\end{equation}
Eq.~(\ref{ampgreen}) and Eq.~(\ref{green}) give us
\begin{equation}
\Lambda_I\pm\Lambda_{\gamma_5}=S^{-1}(p)\left[
\frac{1}{N}\sum_{i=1}^NS_i(p|0)(I\pm\gamma_5)\gamma_5S_i^\dagger(p|0)\gamma_5
\right]S^{-1}(p).
\label{LIpmg5}
\end{equation}
For zero modes with positive chirality, $S_i(p|0)(I-\gamma_5)=0$, while
for zero modes with negative chirality, $S_i(p|0)(I+\gamma_5)=0$.
Therefore, the zero mode contribution in the Green function combination
$G_S(p)\pm G_P(p)$
are removed. We can use Eq.~(\ref{zspeq1}) to obtain a $Z_{SP}$ which
has a suppressed zero mode contribution (Note that $S^{-1}(p)$ still 
contains zero modes).

Similarly, for vector and axial vector currents, Eq.~(\ref{recond}) gives
\begin{equation}
Z_VZ_q\Lambda_{\gamma_\mu}(p^2=\mu^2)=\gamma_\mu,
\end{equation}
and
\begin{equation}
Z_AZ_q\Lambda_{\gamma_\mu\gamma_5}(p^2=\mu^2)=\gamma_\mu\gamma_5.
\end{equation}
Letting $Z_{VA}\equiv Z_V=Z_A$, we obtain
\begin{equation}
\gamma_\mu Z_{VA}Z_q(\Lambda_{\gamma_\mu}\pm\Lambda_{\gamma_\mu\gamma_5})
=I\pm\gamma_5
\end{equation}
and subsequently (after the index $\mu$ is averaged)
\begin{equation}
Z_{VA}Z_q\frac{1}{48}\Tr\left[(\Lambda_{\gamma_\mu}\pm
\Lambda_{\gamma_\mu\gamma_5})\gamma_\mu\right]=1.
\label{zva}
\end{equation}
Similar to Eq.~(\ref{LIpmg5}), we have
\begin{equation}
\Lambda_{\gamma_\mu}\pm\Lambda_{\gamma_\mu\gamma_5}=S^{-1}(p)\left[
\frac{1}{N}\sum_{i=1}^NS_i(p|0)\gamma_\mu(I\pm\gamma_5)
\gamma_5S_i^\dagger(p|0)\gamma_5\right]S^{-1}(p).
\label{Lgmupmgmug5}
\end{equation}
For zero modes with positive chirality, $S_i(p|0)\gamma_\mu(I+\gamma_5)
=S_i(p|0)(I-\gamma_5)\gamma_\mu=0$. For those with negative chirality,
$S_i(p|0)\gamma_\mu(I-\gamma_5)=S_i(p|0)(I+\gamma_5)\gamma_\mu=0$.
Therefore, $Z_{VA}$ calculated from Eq.~(\ref{zva}) has a suppressed zero
mode contribution. 

\section{Numerical Results}\label{NumRes}
The data set that we use contains 40 gauge configurations in the
quenched approximation with the Wilson gauge action. The lattice size
is $16^4$ and the gauge coupling $\beta=6.1$. The bare quark masses in
lattice units are $am_q=0.015$, 0.020, 0.025, 0.035, 0.050 and 0.070.
The lattice spacing $a$ is $(a)$:~0.08~fm determined from the 
interpolation formula of
Ref.~\cite{sommer} using the Sommer parameter or $(b)$:~0.09~fm from the
measured rho mass. Therefore, $a\mu=0.811$ or $a\mu=0.913$
corresponds to $\mu=2$~GeV accordingly.
In the following analysis,
the statistical errors are obtained by a Jackknife average with one
configuration removed each time.

\subsection{$Z_q^{RI'}$ and $Z_m^{RI'}$}
\begin{figure}
{\centering\includegraphics[width=80mm]
{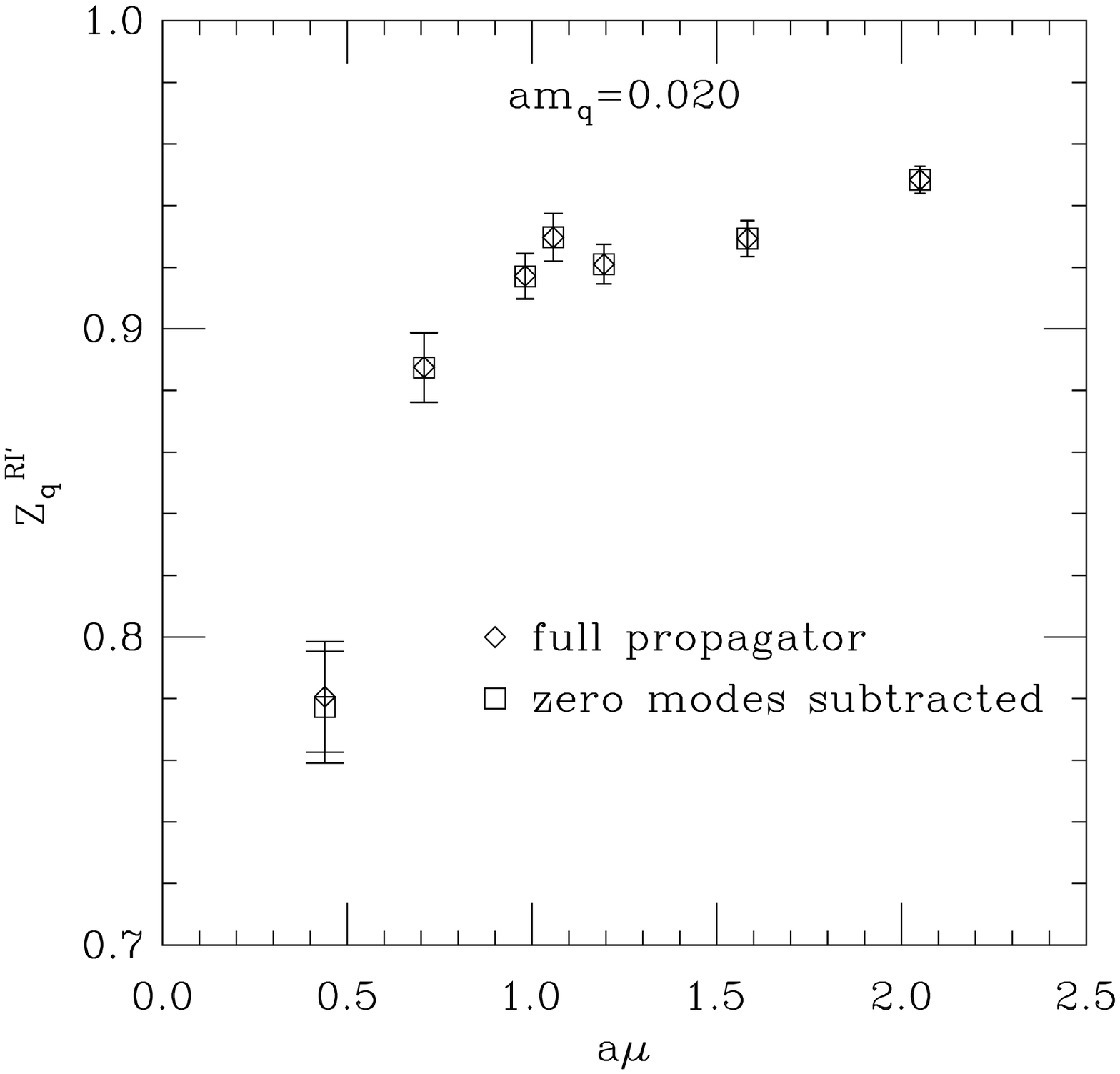}
\includegraphics[width=80mm]
{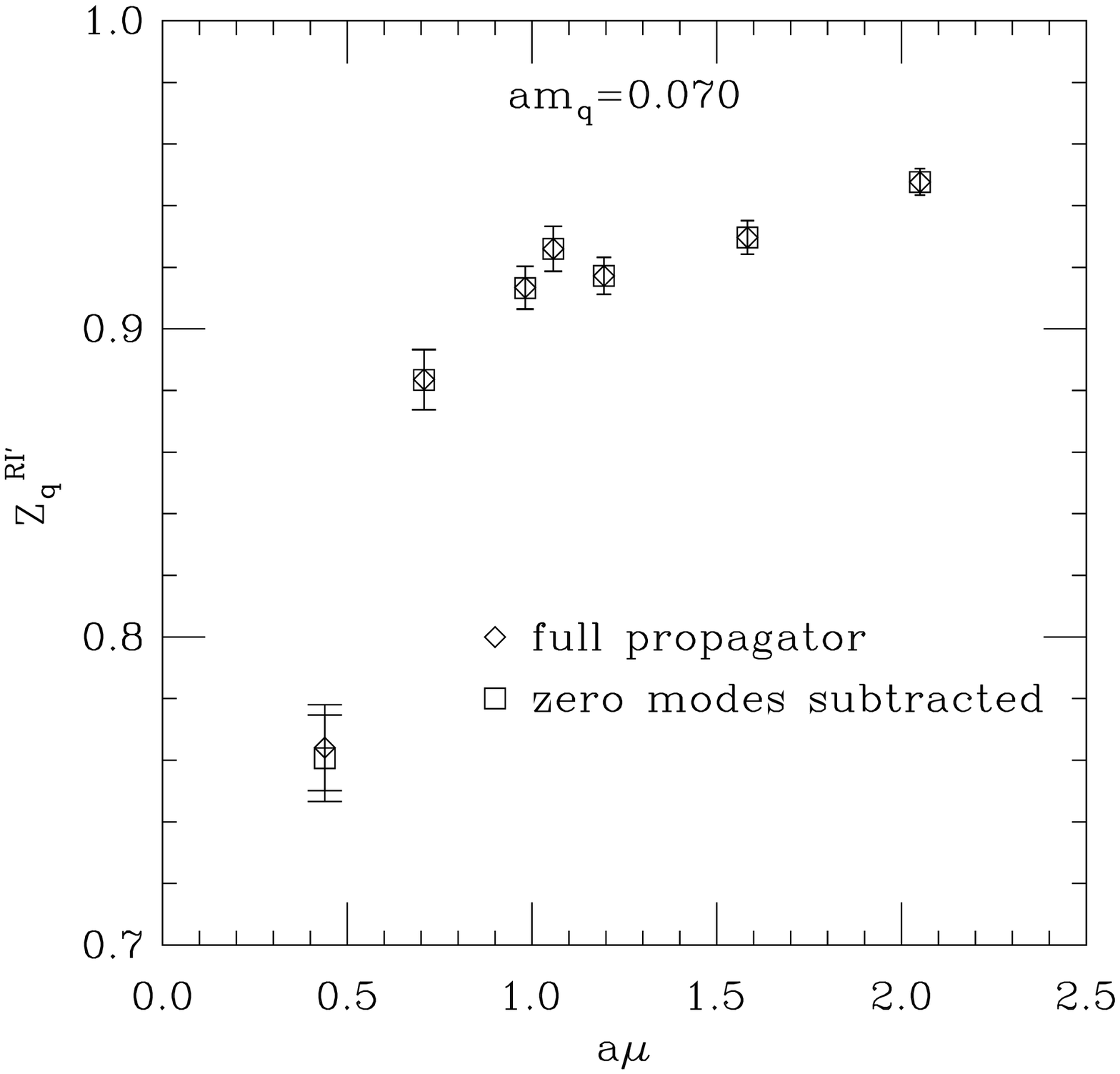}}
\caption{$Z_q^{RI'}$ vs. $a\mu$ for bare quark mass $am_q=0.020$
and 0.070. $Z_q^{RI'}$ obtained from the full propagator (diamond)
is compared with that obtained from the propagator with zero mode
contribution subtracted (square).}
\label{zqfig1}
\end{figure}

The quark field renormalization constant $Z_q^{RI'}$ is calculated with
Eq.~(\ref{zqfactor}). The results for two examples of 
bare quark masses are shown
in Fig.~\ref{zqfig1}. The comparison between $Z_q^{RI'}$ obtained from
the full propagator and the propagator with zero mode subtracted is also
shown in the same graph. There is no difference within error bars.

\begin{figure}
{\centering\includegraphics[width=80mm]
{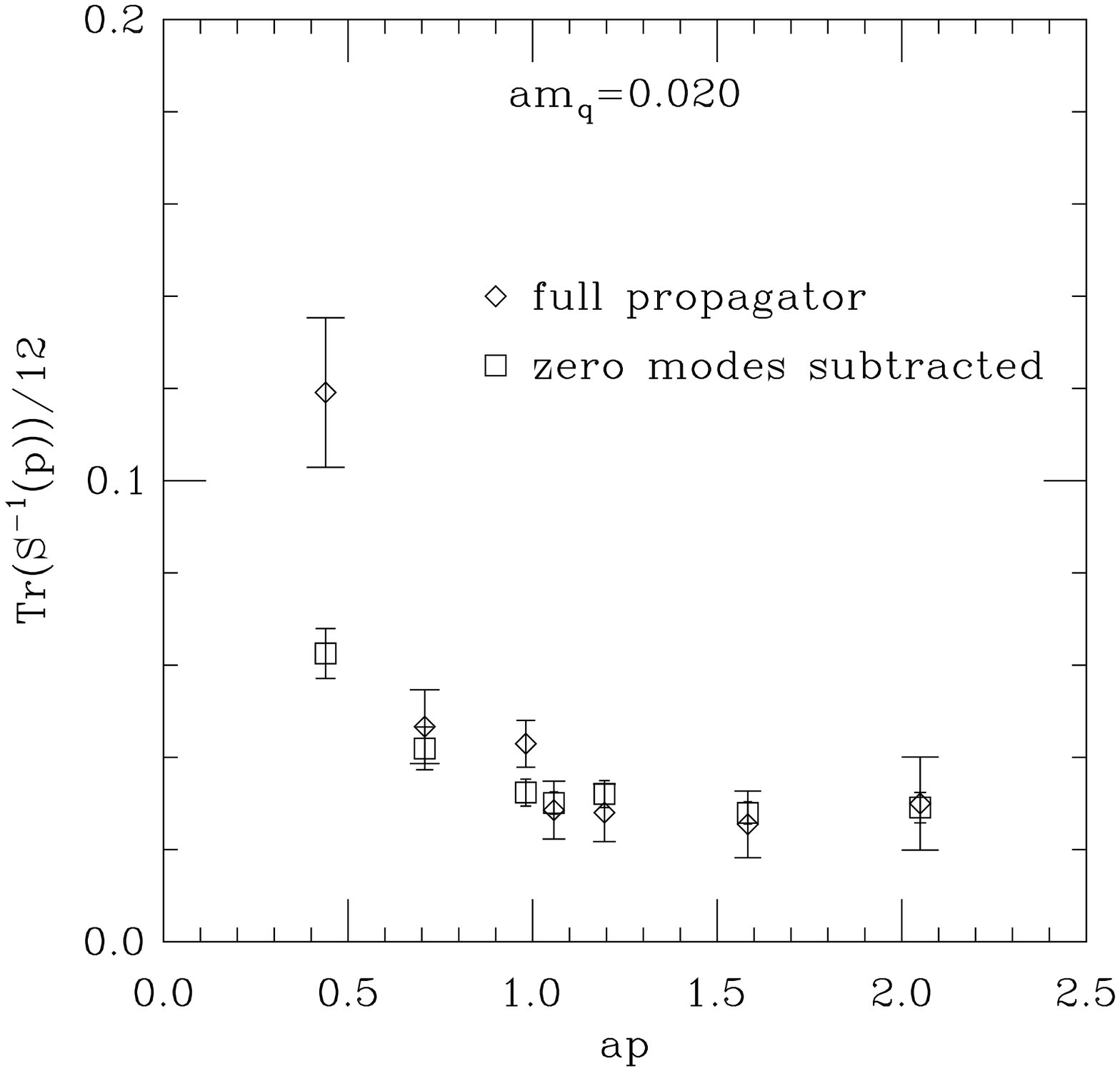}
\includegraphics[width=80mm]
{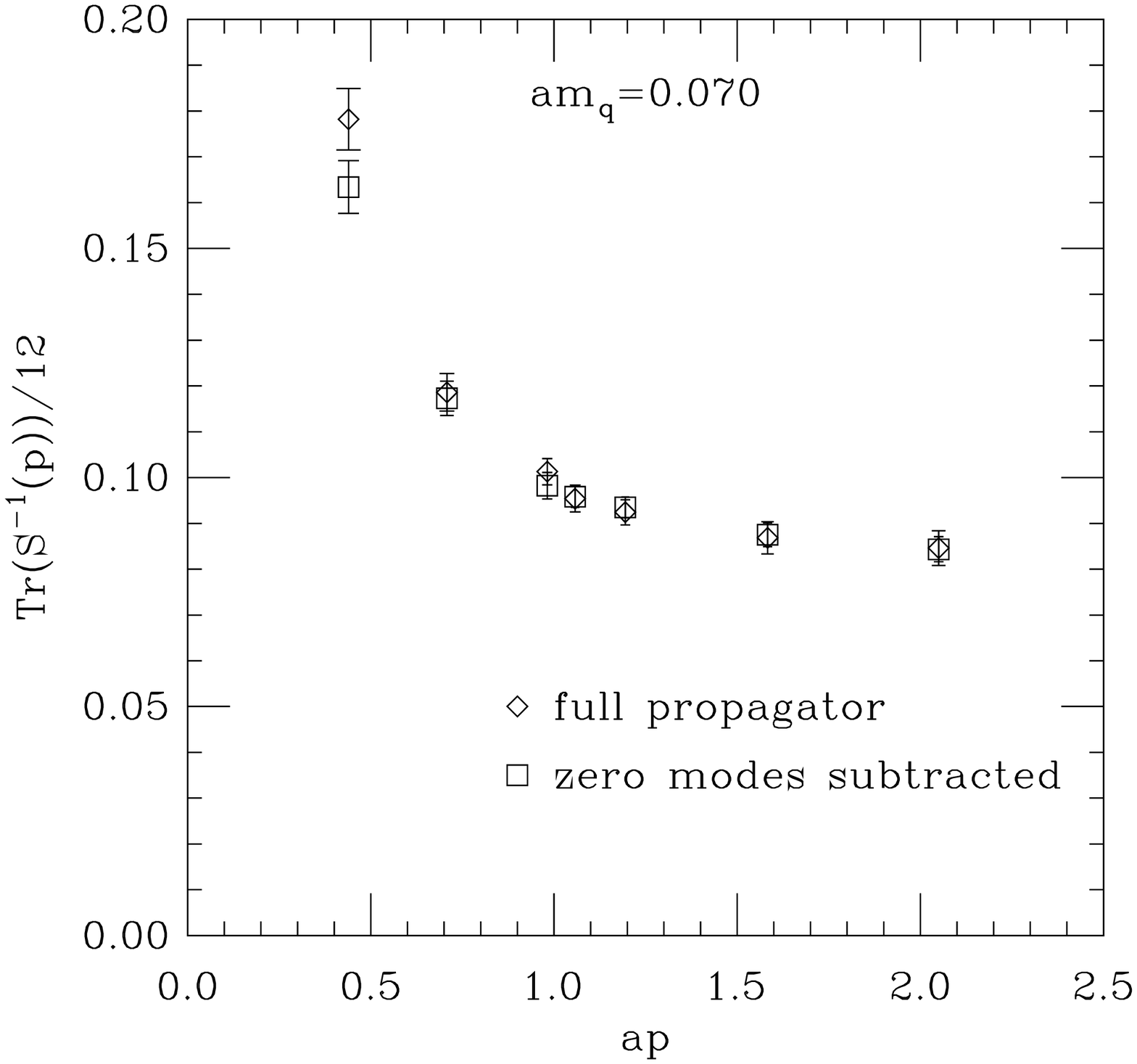}}
\caption{$(1/12)\Tr(S^{-1}(p))$ vs. $ap$ for bare quark masses
$am_q=0.020$ and 0.070. Results from the full propagator (diamond)
and from the propagator with zero mode contribution subtracted (square)
are compared. Zero mode contribution is important only at small momentum
and small quark mass.}
\label{aMq}
\end{figure}

The full lattice quark propagator takes the form
\begin{equation}
S(p)=\frac{Z(p)}{i\gamma\cdot q(p)+M(p)}.
\label{flp}
\end{equation}
Here $q(p)$ is the kinematic momentum depending on the lattice quark
action one uses.
At large momentum $p$, because of asymptotic freedom the propagator
should go back to the free quark propagator. i.e. $Z(p)\rightarrow1$
and $M(p)$ goes to the bare quark mass.
Fig.~\ref{aMq} shows
$(1/12)\Tr(S^{-1}(p))$ versus
$ap$ for two examples of bare quark masses with $S(p)$ determined
from Eq.~(\ref{sp}). Results from the full propagator and
from the propagator with zero mode contribution subtracted
are compared in the graph.
As is expected, $(1/12)\Tr(S^{-1}(p))$ approaches the bare quark mass at
large momentum. Apparently, only at small momentum and small quark mass
does the zero mode contribution make a difference.

\begin{figure}
{\centering\includegraphics[width=80mm]
{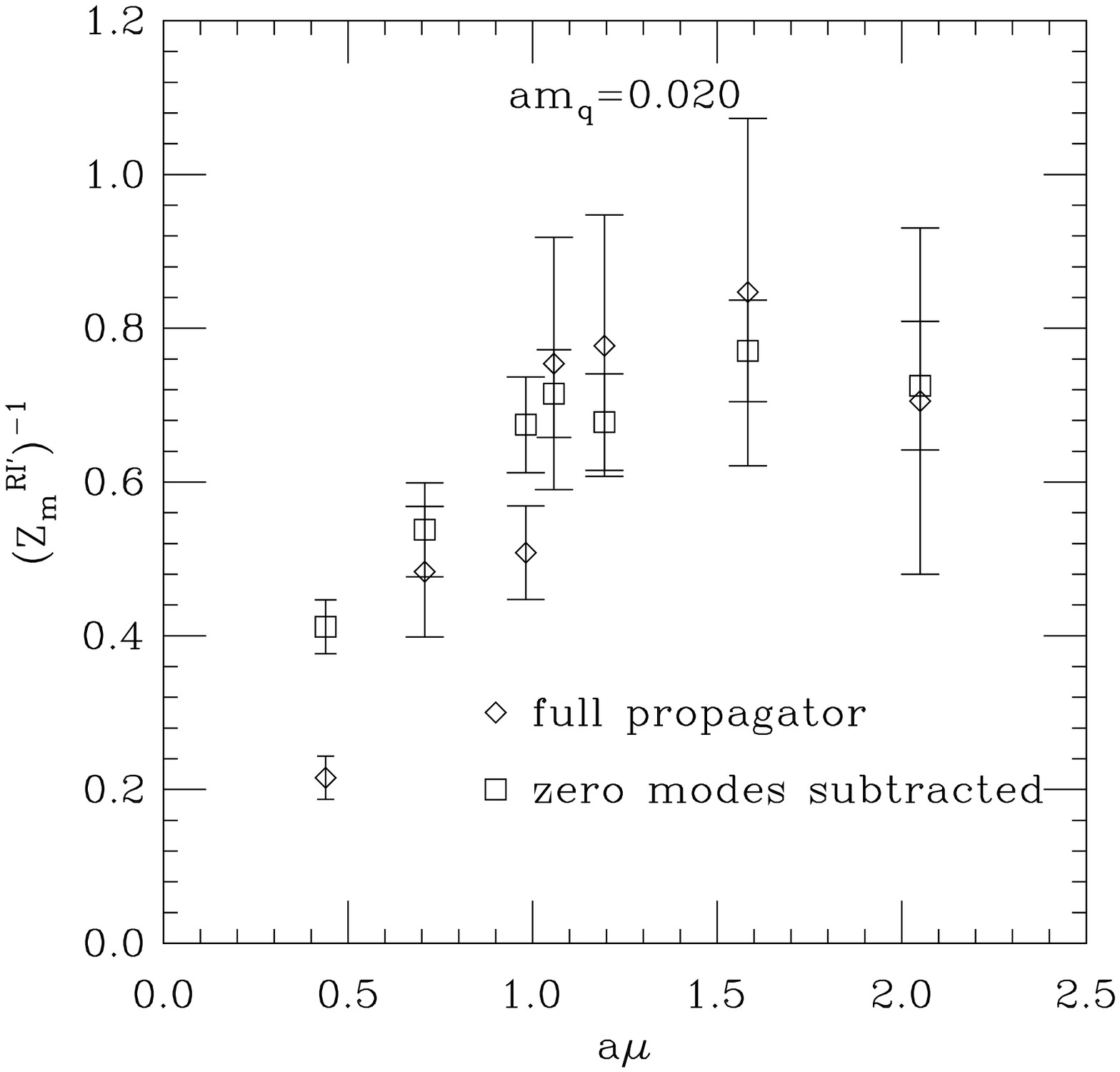}
\includegraphics[width=80mm]
{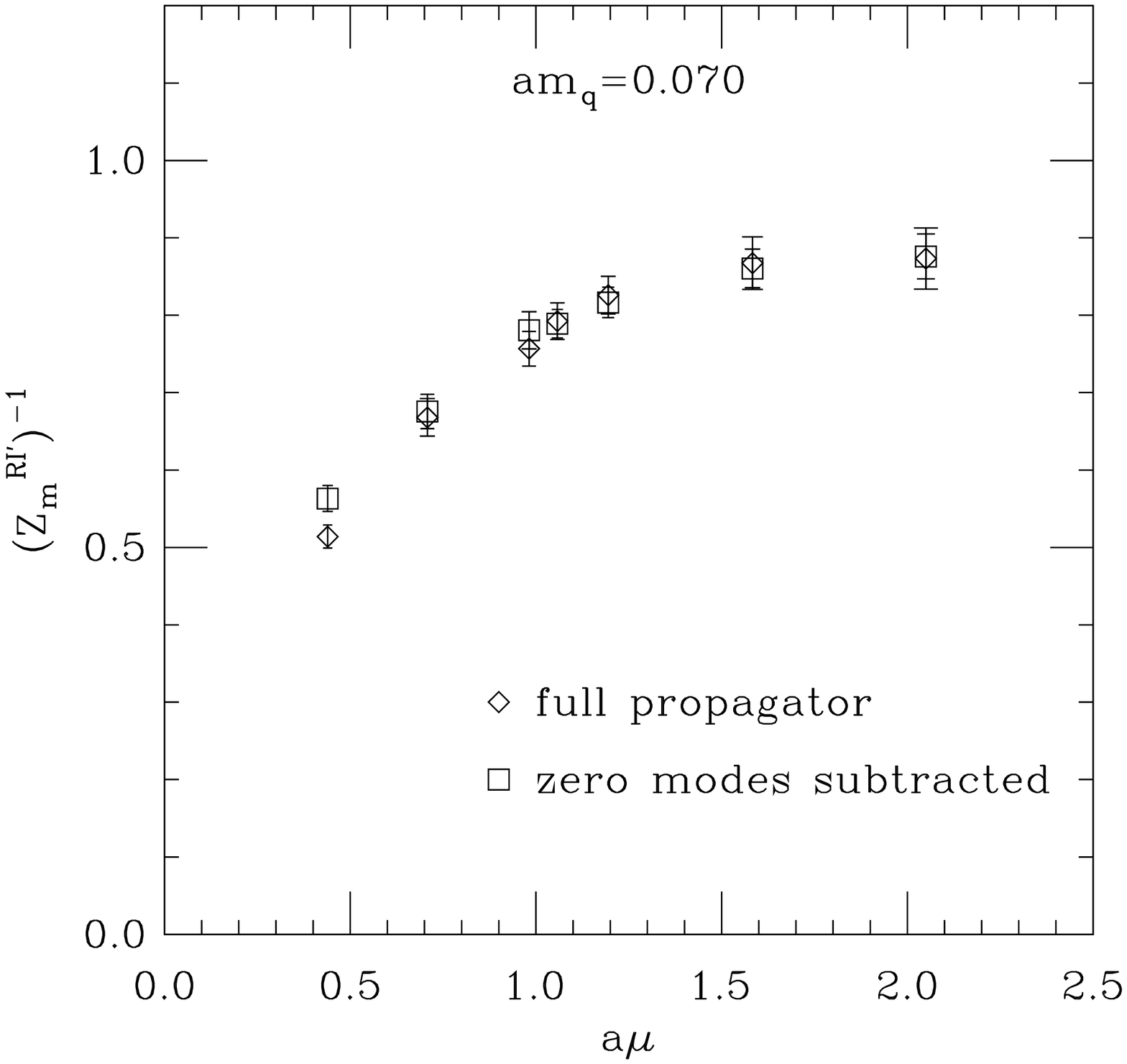}}
\caption{$(Z_m^{RI'})^{-1}$ for bare quark masses
$am_q=0.020$ and 0.070. Diamonds are from the full propagator
and squares from the propagator with zero mode contribution subtracted.}
\label{zm1}
\end{figure}

If we define a renormalized quark mass $m(\mu)$ by
\begin{equation}
m(\mu)=Z_m(\mu)m_0,
\end{equation}
where $m_0$ is the bare quark mass, then $Z_m(\mu)$ is fixed in
the RI' scheme by
\begin{equation}
(Z_m^{RI'})^{-1}=\lim_{m\rightarrow 0}\left.\frac{12m_0}{Z_q^{RI'}\Tr(S^{-1}(p))}
\right|_{p^2=\mu^2}.
\end{equation}
At finite quark masses, the renormalization conditions of RI' scheme are
compatible with the Ward identities~\cite{npr1,fran} at large $\mu^2$, 
therefore we expect
$Z_m^{RI'}=Z_S^{-1}$ at large $\mu^2$.
The numerical results of $(Z_m^{RI'})^{-1}$ are shown in Fig.~\ref{zm1}.
The error bar at small quark mass is large.
A zero mode contribution is visible only at low momentum.
We will compare $(Z_m^{RI'})^{-1}$ with $Z_S$ later.

\subsection{$Z_S^{RI'}$ and $Z_P^{RI'}$}
\begin{figure}
{\centering\includegraphics[width=80mm]
{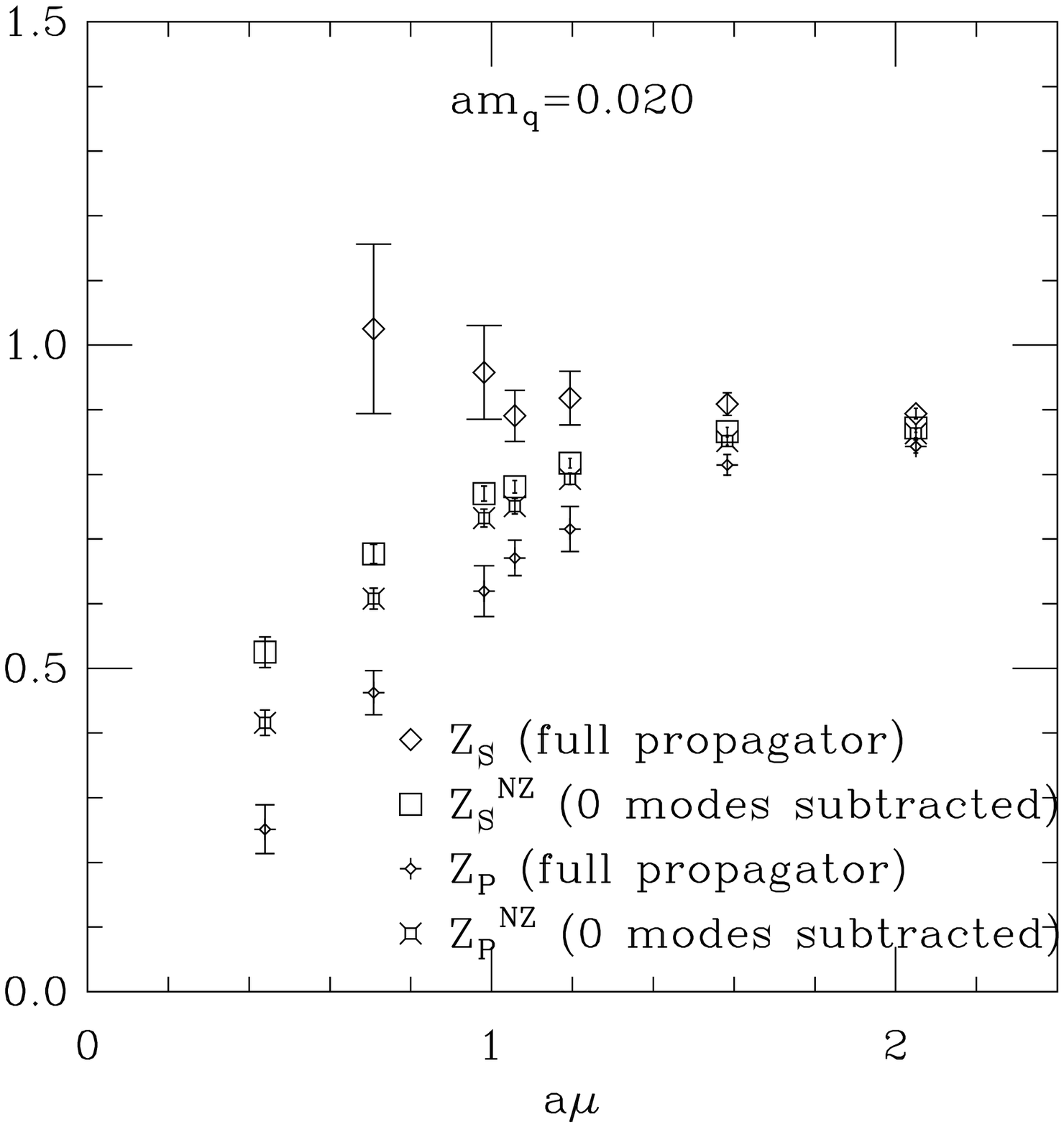}
\includegraphics[width=80mm]
{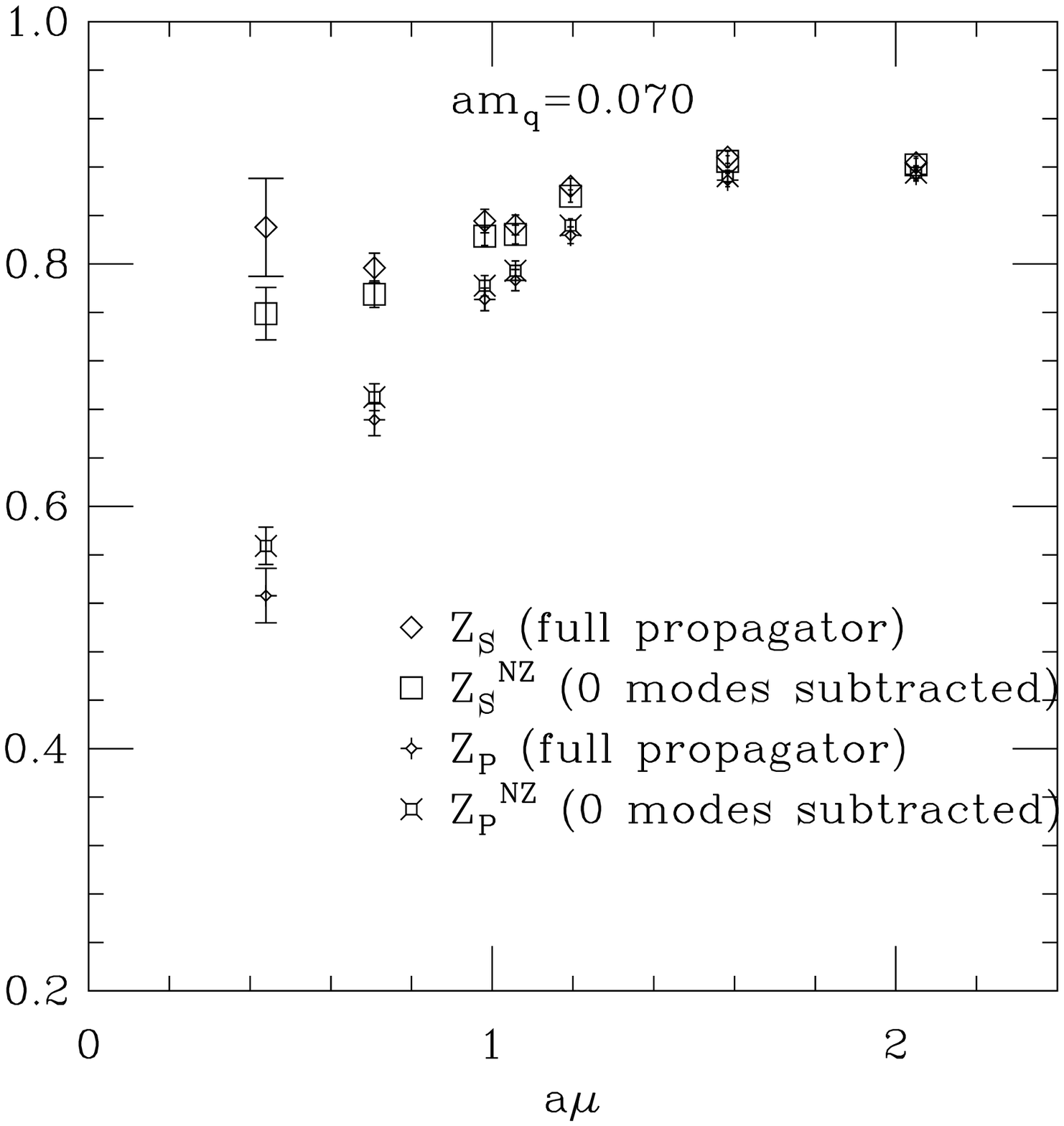}}
\caption{$Z_S$ and $Z_P$ from the full propagators and the zero modes
subtracted (labeled as $Z_S^{NZ}$ and $Z_P^{NZ}$) propagators. 
The left is for quark mass $am_q=0.020$, the
right $am_q=0.070$.}
\label{zsandzp}
\end{figure}

The results for $Z_S$ and $Z_P$ are shown in Fig.~\ref{zsandzp}.
At low quark mass and momentum region, the zero mode subtracted
propagators give different values of $Z_S$ and $Z_P$ (we will label them
as $Z_S^{NZ}$ and $Z_P^{NZ}$ in the following)
from those obtained with the full propagators. The pseudoscalar
density couples to the Goldstone boson channel but the coupling is
suppressed at large $\mu$~\cite{npr1}. Therefore we see a
difference between $Z_S$ and $Z_P$ at small $\mu$, but no difference
at large $\mu$.

\begin{figure}
{\centering\includegraphics[width=80mm]
{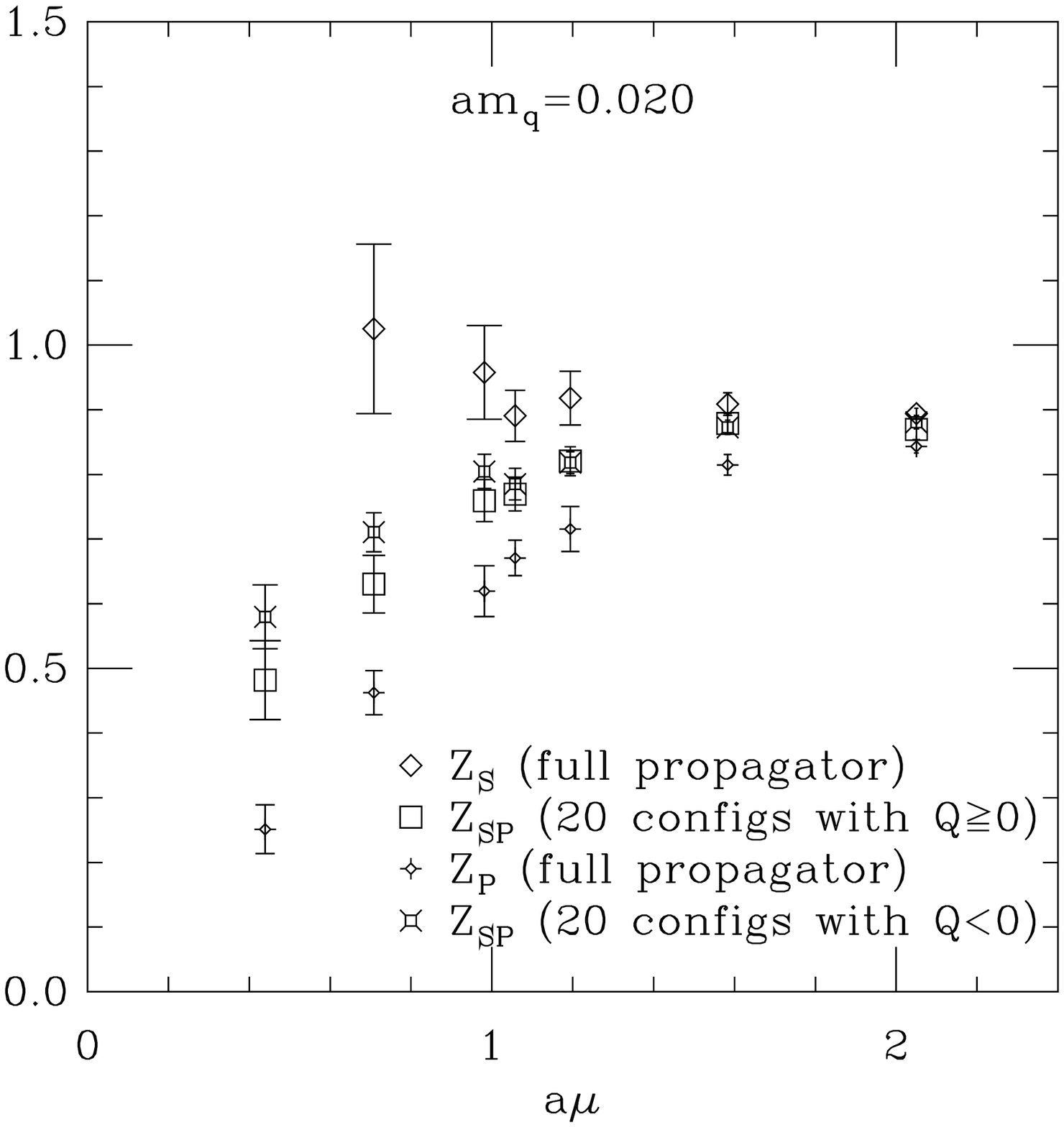}
\includegraphics[width=80mm]
{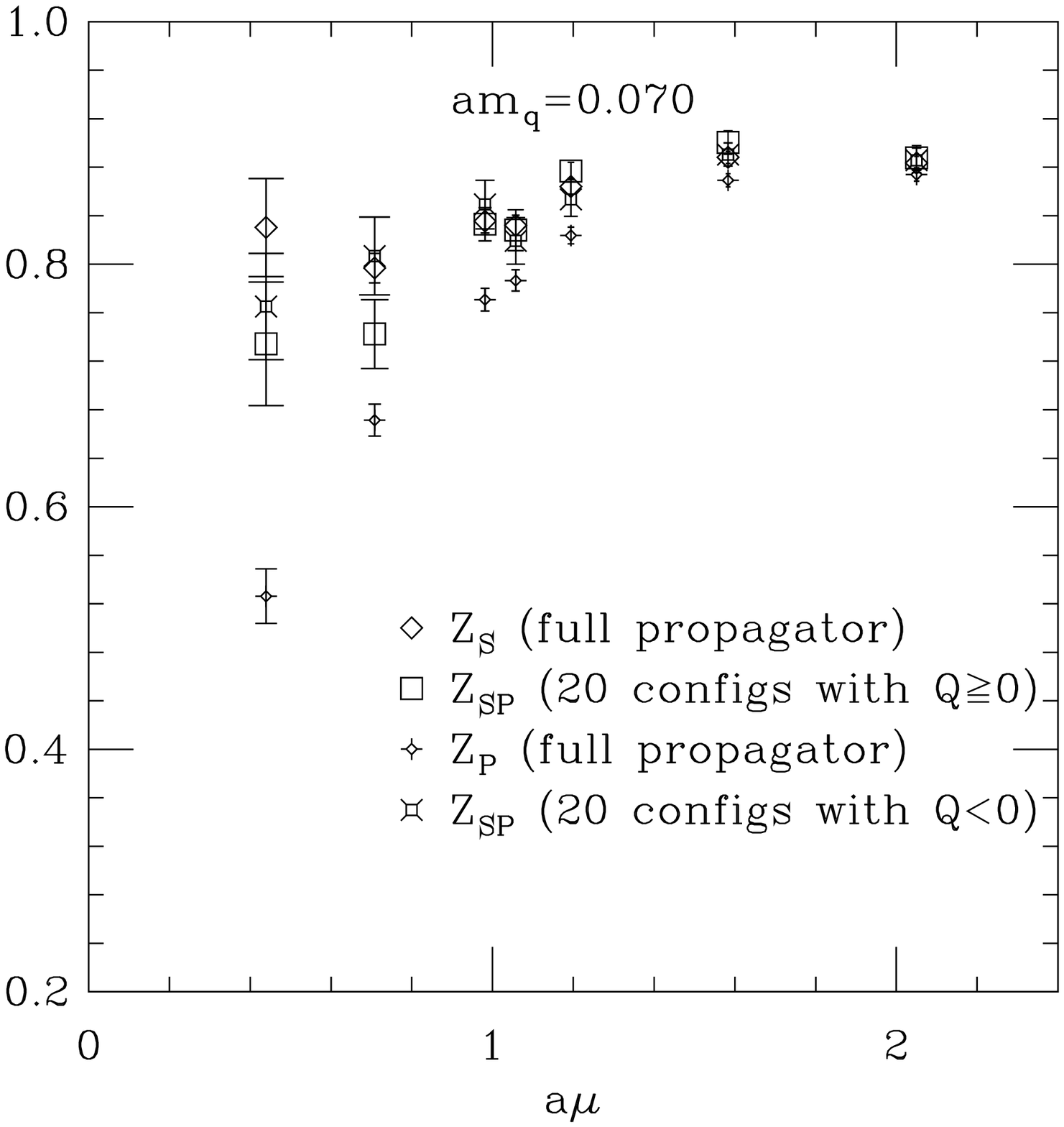}}
\caption{$Z_{SP}$ obtained by using Eq.~(\ref{zspeq1}) comparing with 
$Z_S$ and $Z_P$. The left is for quark mass $am_q=0.020$, the
right $am_q=0.070$. $Q$ is the topological charge.}
\label{zszpzsp1}
\end{figure}
As was discussed in Section \ref{Meth}, we can also use Eq.~(\ref{zspeq1}) to 
suppress the zero modes and obtain $Z_{SP}$. Fig.~\ref{zszpzsp1} shows
the results of $Z_{SP}$ comparing with $Z_S$ and $Z_P$. The $Z_{SP}$'s
from configurations with topological charge $Q\ge0$ and $Q<0$ are
quite close to each other. At small quark mass and small $\mu$,
$Z_{SP}$ is apparently different from $Z_S$ and $Z_P$ as can be seen
in the graph for $am_q=0.020$ in Fig.~\ref{zszpzsp1}. $Z_{SP}$ agrees
with $Z_S^{NZ}$ and $Z_P^{NZ}$ obtained from the zero mode 
subtracted propagators
in Fig.~\ref{zsandzp} for $am_q=0.020$. For large quark mass, for
example $am_q=0.070$, and small $\mu$, $Z_{SP}$ is close to $Z_S$ but
very different from $Z_P$. $Z_{SP}$ from Eq.~(\ref{zspeq1}) is still
contaminated by the coupling to the Goldstone boson. Thus, this means
that suppressing the zero modes can suppress the coupling to the
Goldstone boson. We do not see this behavior in Fig.~\ref{zsandzp}
for $am_q=0.070$. The $Z_P^{NZ}$ from the zero mode subtracted
propagator is very close to the $Z_P$ from the full propagator.
However, we should notice that
subtracting zero modes directly from 
the propagators amounts to a modification of the quenched theory.
To further investigate the zero modes, a quenched artifact, simulations with
dynamical fermions are necessary.

\begin{figure}
{\centering\includegraphics[width=80mm]
{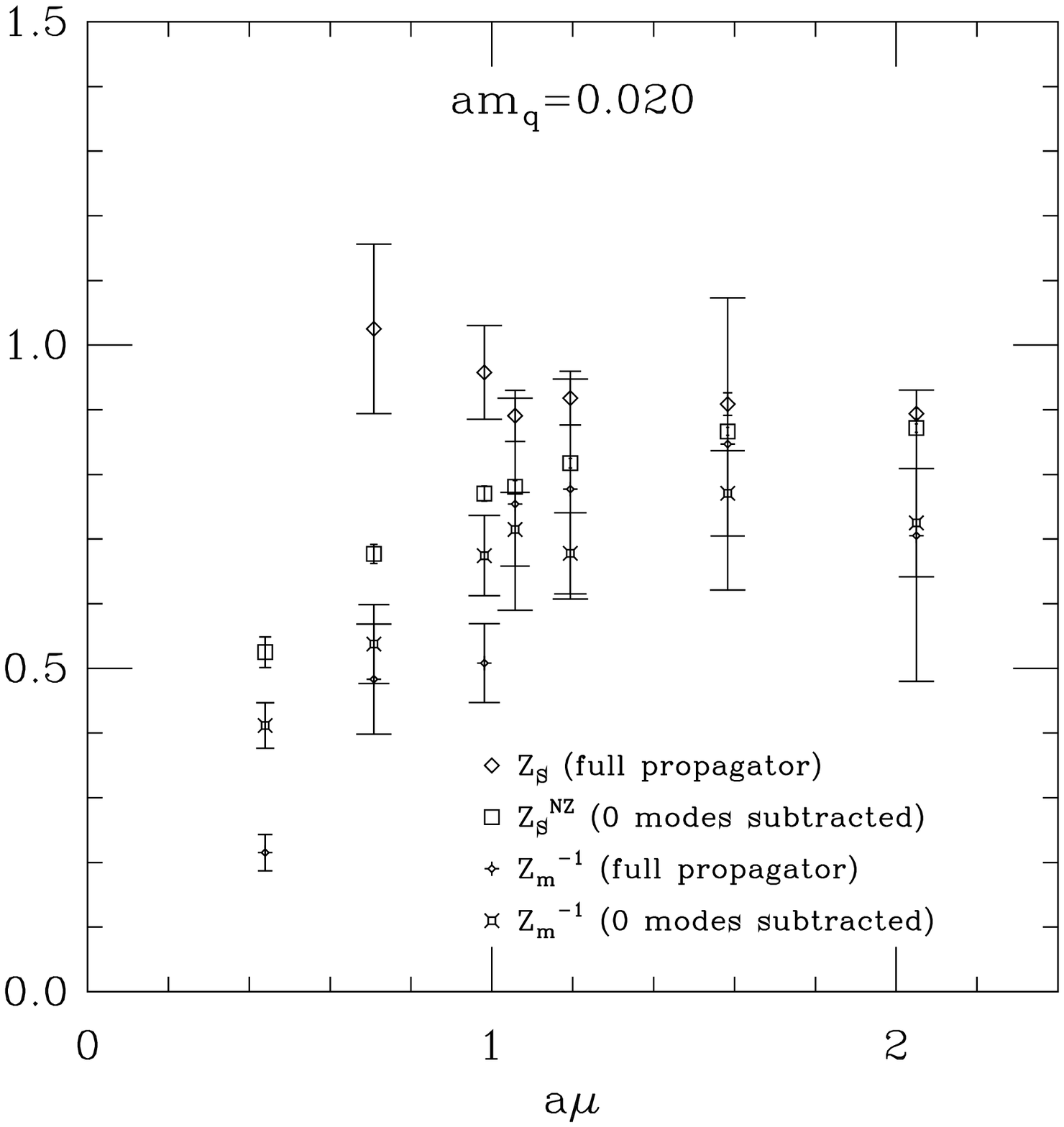}
\includegraphics[width=80mm]
{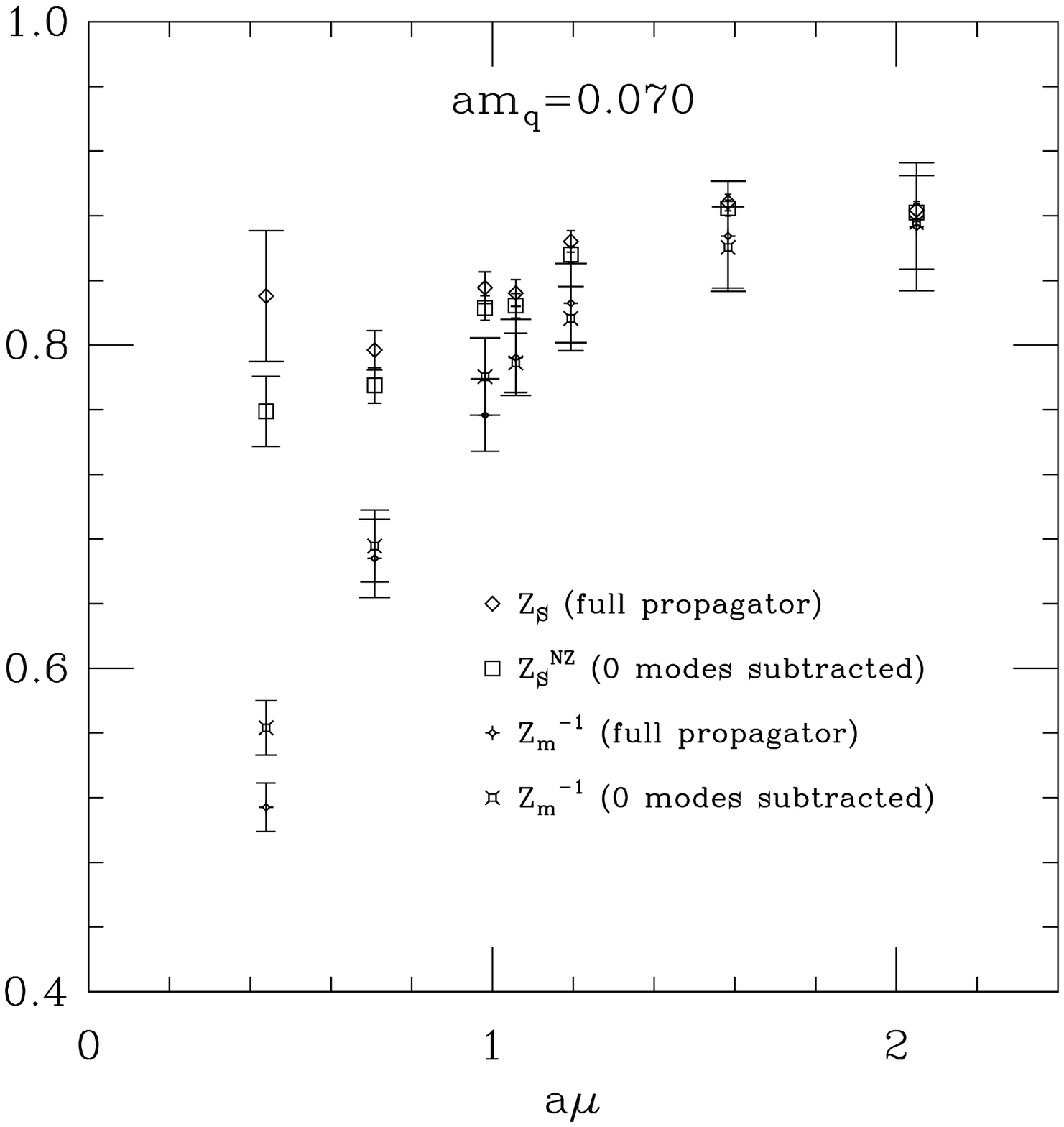}}
\caption{$(Z_m^{RI'})^{-1}$ compared with $Z_S^{RI'}$. At large $\mu$,
$(Z_m^{RI'})^{-1}=Z_S^{RI'}$ is well satisfied.}
\label{zszm}
\end{figure}
The comparison of $(Z_m^{RI'})^{-1}$ with $Z_S^{RI'}$ is given in
Fig.~\ref{zszm}. We see a good agreement between $(Z_m^{RI'})^{-1}$ and
$Z_S^{RI'}$ at large $\mu$ as expected.

\begin{figure}
{\centering\includegraphics[width=80mm]
{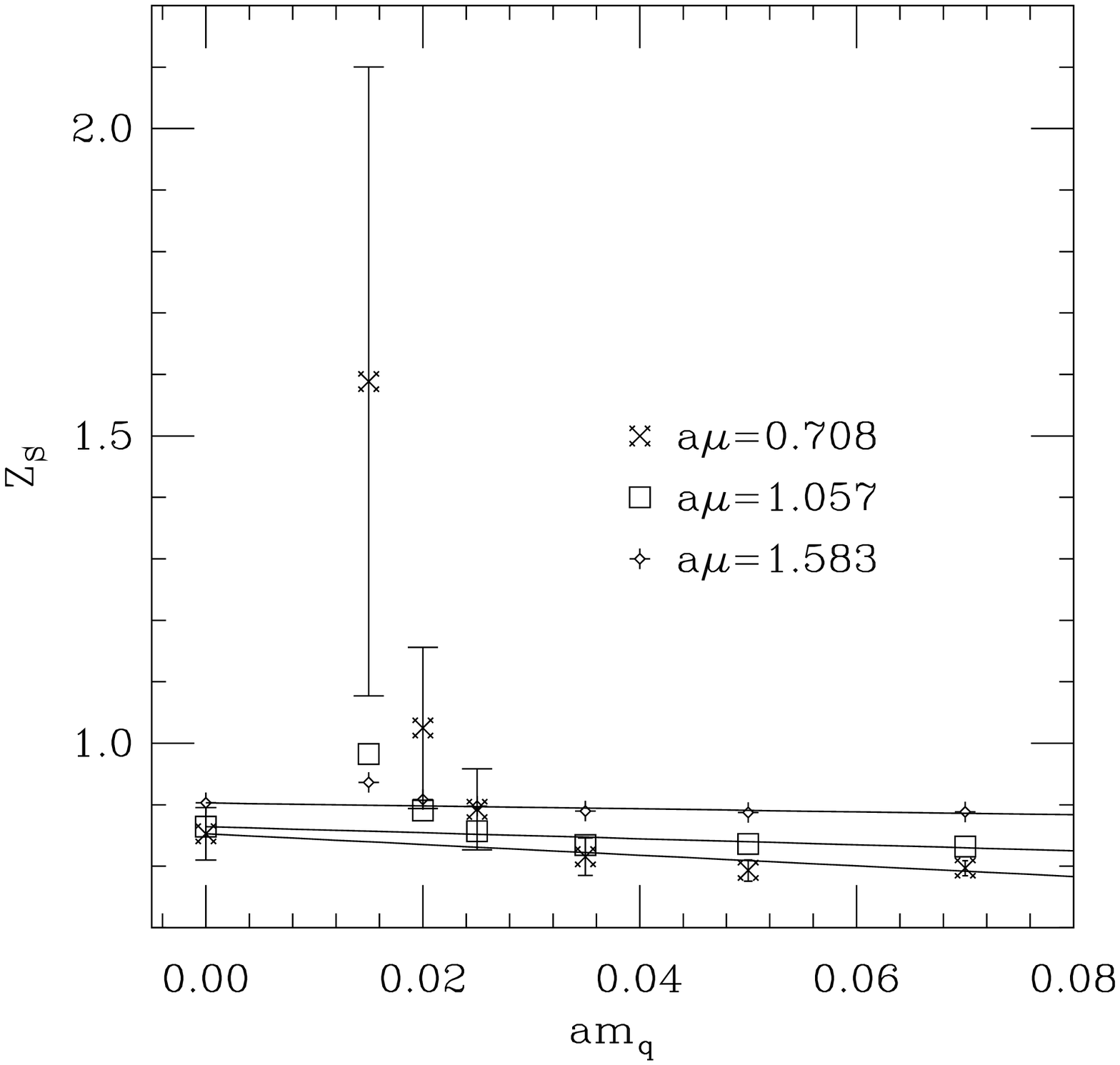}
\includegraphics[width=80mm]
{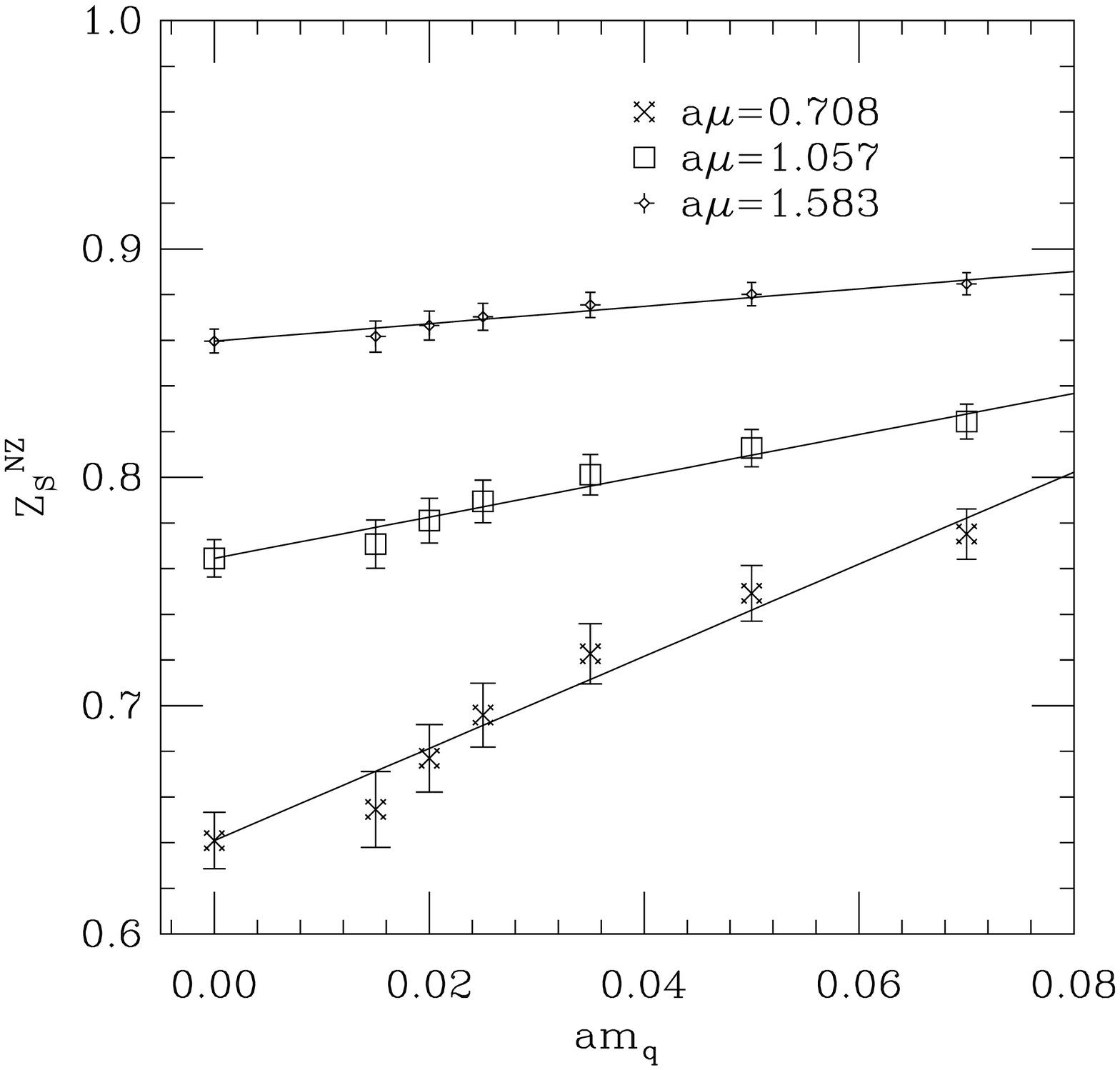}}
\caption{Linear extrapolation of $Z_S$ to the chiral limit.
The left is for $Z_S$ obtained from full propagators. 
The data do not support a linear extrapolation. We did it
anyway just for comparison. The right
is for $Z_S$ obtained from zero mode subtracted propagators
(labeled as $Z_S^{NZ}$). At small quark mass, $Z_S$ and $Z_S^{NZ}$
are very different.}
\label{zsext1}
\end{figure}
\begin{figure}
{\centering\includegraphics[width=80mm]
{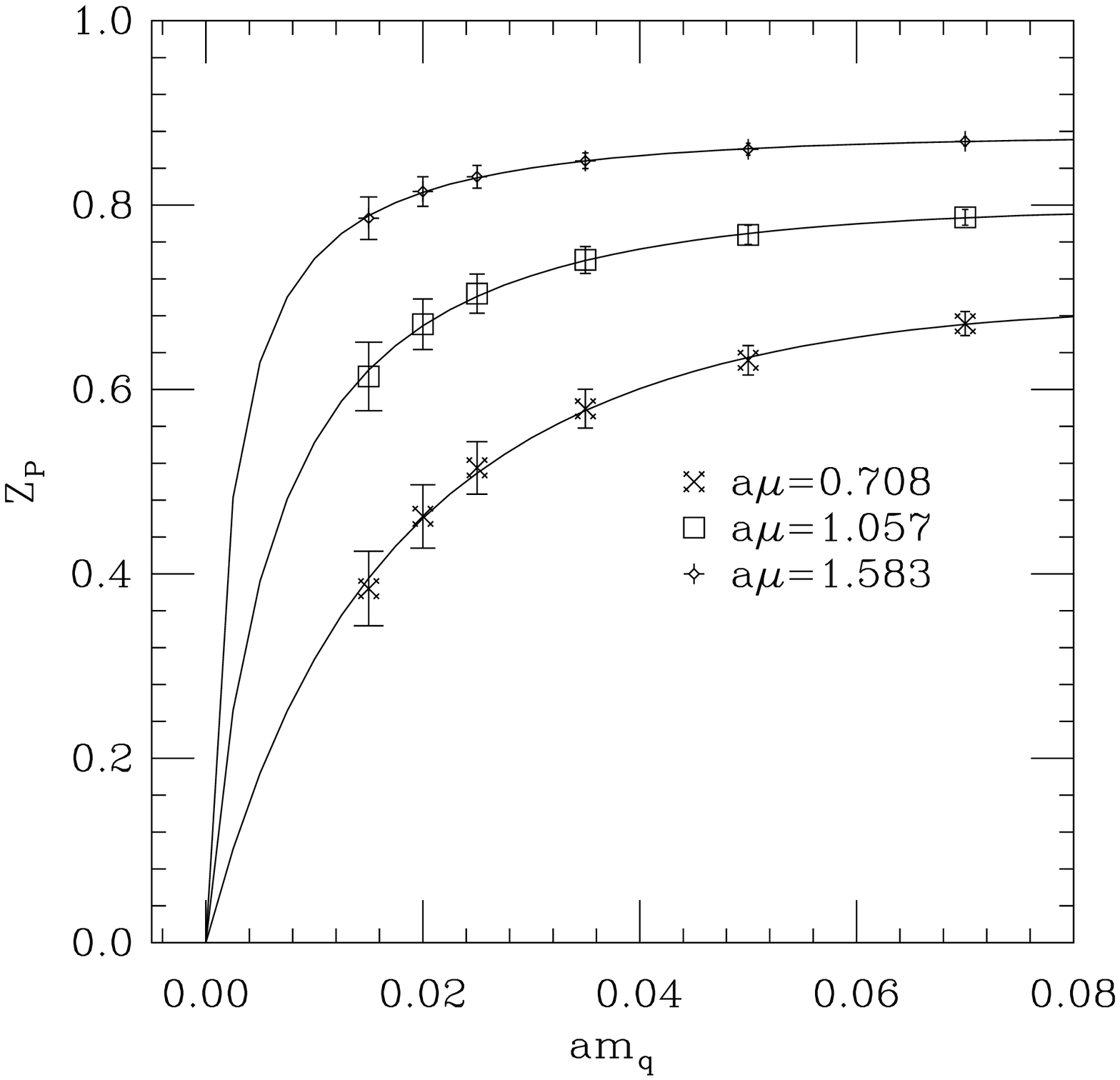}
\includegraphics[width=80mm]
{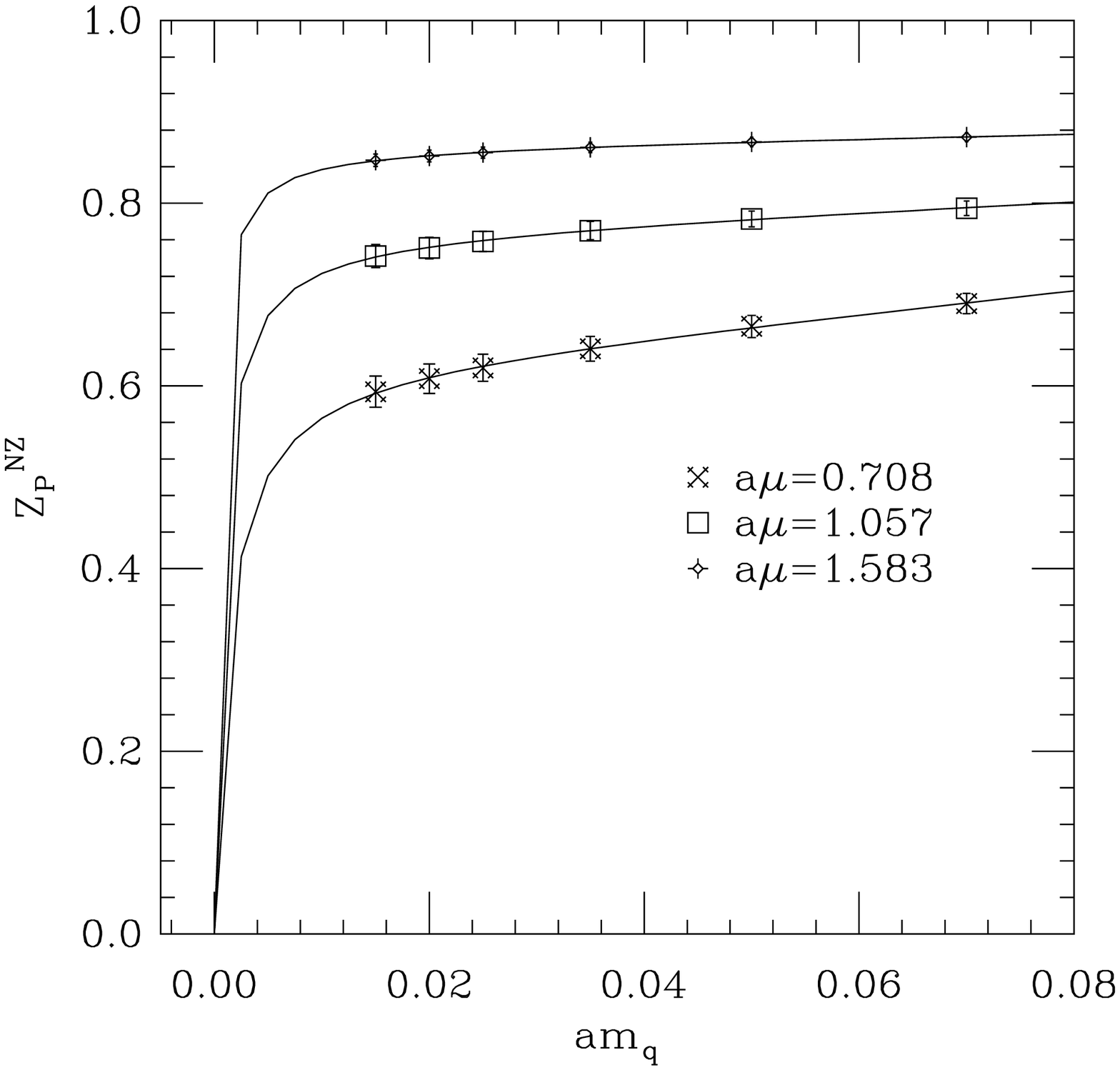}}
\caption{Extrapolation of $Z_P$ to the chiral limit using
Eq.~(\ref{zpextrp}). $Z_P$ in the
left graph is obtained by using the full propagator, while $Z_P^{NZ}$
in the right graph is obtained by using the propagator with zero modes
subtracted. Both are fits after the last configuration is dropped during
the Jackknife average process.}
\label{zpexp}
\end{figure}
\begin{figure}
{\centering\includegraphics[width=80mm]
{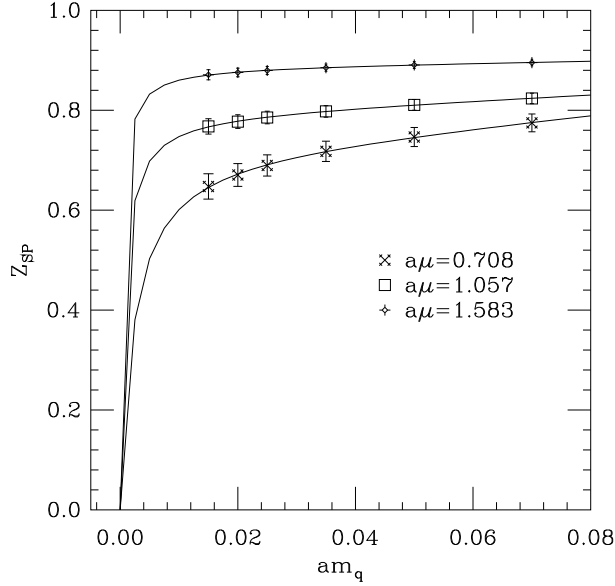}
\caption{Extrapolation of $Z_{SP}$ to the chiral limit using
Eq.~(\ref{zpextrp}). $Z_{SP}$ is the average from configurations
with $Q\ge0$ and $Q<0$.}
\label{zspextrp}}
\end{figure}

In Fig.~\ref{zsext1}, \ref{zpexp} and \ref{zspextrp}, $Z_S$, $Z_S^{NZ}$,
$Z_P$, $Z_P^{NZ}$ and $Z_{SP}$(average from configurations with  $Q\ge0$ 
and $Q<0$) are plotted
versus the quark mass at $a\mu=0.708$, $1.057$ and $1.583$ along with
the extrapolation to the chiral limit.
A linear fit is used for $Z_S$ and $Z_S^{NZ}$. 
The data for $Z_S$ which is obtained from the
full propagator do not support a linear extrapolation as is shown in
Fig.~\ref{zsext1}.
For $Z_P$, we see similar behavior as was seen 
in \cite{gock,gatt,cude,gius,beci}
since the pseudoscalar density couples to the Goldstone boson channel.
As in \cite{gatt,beci}, we use
\begin{equation}
\frac{1}{Z_P(\mu^2,m)}=\frac{A(\mu^2)}{am_q}+B(\mu^2)+C(\mu^2)(am_q)
\label{zpextrp}
\end{equation}
to fit $Z_P$ and then remove the pole term $A(\mu^2)/am_q$
to obtain $Z_P^{NP}\equiv B(\mu^2)^{-1}$ in the chiral limit. The
fit is good as can be seen in Fig.~\ref{zpexp}. Fig.~\ref{zpexp} shows
one example of the fit when we drop one configuration and obtain
$Z_P$ with the rest configurations during the Jackknife average process.

We also use Eq.~(\ref{zpextrp}) to extrapolate $Z_{SP}$ to the chiral
limit to obtain $Z_{SP}^{NP}\equiv B(\mu^2)^{-1}$. The fit
is shown in Fig.~\ref{zspextrp}.

Values of $Z_S$, $Z_P^{NP}$, $Z_S^{NZ}$, $(Z_P^{NZ})^{NP}$ and
$Z_{SP}^{NP}$ in the RI' scheme 
in the chiral limit
are listed in Table \ref{zszpsubtb1}. In the table, the superscripts
$(a)$ and $(b)$ indicate the two different ways of determining the
lattice spacing. We could not get $Z_S$ 
for the smallest momentum because the signal is noisy in our data.
$Z_S$ and $Z_P^{NP}$ are different from each other at small $\mu$ so
that the $\overline{MS}$ values of them differ from each other.
The $\overline{MS}$ values are obtained by using the conversion
formulas in Section \ref{Conv} and a linear interpolation from the
two closest $\mu$ value of the data. 
After subtracting the zero modes from the
quark propagators, we get $Z_S^{NZ}$ and $(Z_P^{NZ})^{NP}$. They
are in good agreement with each other, which is expected from chiral
symmetry, but are very different from $Z_S$ and $Z_P^{NP}$.
The other way of suppressing the zero modes gives us $Z_{SP}^{NP}$.
Since the zero mode effects are much more apparent in $Z_S$ and $Z_P$,
we will
use $Z_S^{NZ}$, $(Z_P^{NZ})^{NP}$ and $Z_{SP}^{NP}$ to compare with
perturbative calculations.
\begin{table}
\begin{center}
\begin{tabular}{|c|c|c|c|c||c|c|c|}
$a\mu$&$\mu^{(a)}$(GeV)&$\mu^{(b)}$(GeV)&$Z_S$&$Z_P^{NP}$&$Z_S^{NZ}$&
$(Z_P^{NZ})^{NP}$ & $Z_{SP}^{NP}$ \\
0.439&1.08&0.96&        &1.31(6) &0.43(2) &0.455(2) &0.642(2) \\
0.708&1.75&1.55&0.85(4) &0.97(1) &0.64(1) &0.631(1) &0.732(2) \\
0.982&2.42&2.15&0.89(3) &0.94(2) &0.753(9)&0.755(1) &0.828(2) \\
1.057&2.61&2.32&0.86(2) &0.887(9)&0.764(8)&0.7685(9)&0.7972(9)\\
1.194&2.95&2.62&0.89(2) &0.92(1) &0.805(5)&0.817(1) &0.8489(9)\\
1.583&3.91&3.47&0.90(1) &0.906(4)&0.860(5)&0.8611(3)&0.8874(6)\\
2.050&5.06&4.49&0.896(7)&0.901(2)&0.868(6)&0.8725(2)&0.8826(4)\\
\hline\hline
$\overline{MS}^{(a)}$&2~GeV&&1.01(2)&1.12(1)&0.79(5)&0.79(5)&0.89(4)\\
$\overline{MS}^{(b)}$&&2~GeV&1.01(1)&1.091(9)&0.83(3)&0.83(4)&0.93(3)
\end{tabular}
\end{center}
\caption{Values of $Z_S$, $Z_P^{NP}$, $Z_S^{NZ}$, $(Z_P^{NZ})^{NP}$ and
$Z_{SP}^{NP}$ in the RI' scheme
in the chiral limit. The $\overline{MS}$(2~GeV) value is obtained from
a linear interpolation from the two closest $\mu$ values of the data.
The lattice spacing is $(a)$:~0.08~fm from Sommer parameter or 
$(b)$:~0.09~fm from the measured rho mass. Correspondingly, we get
two $\overline{MS}$(2~GeV) values.
$Z_S$ and $Z_P^{NP}$ contains zero modes and seen to be quite different
from the $Z$'s which contains no zero modes.}
\label{zszpsubtb1}
\end{table}

\subsection{$Z_V^{RI'}$ and $Z_A^{RI'}$}
\begin{figure}
{\centering\includegraphics[width=80mm]
{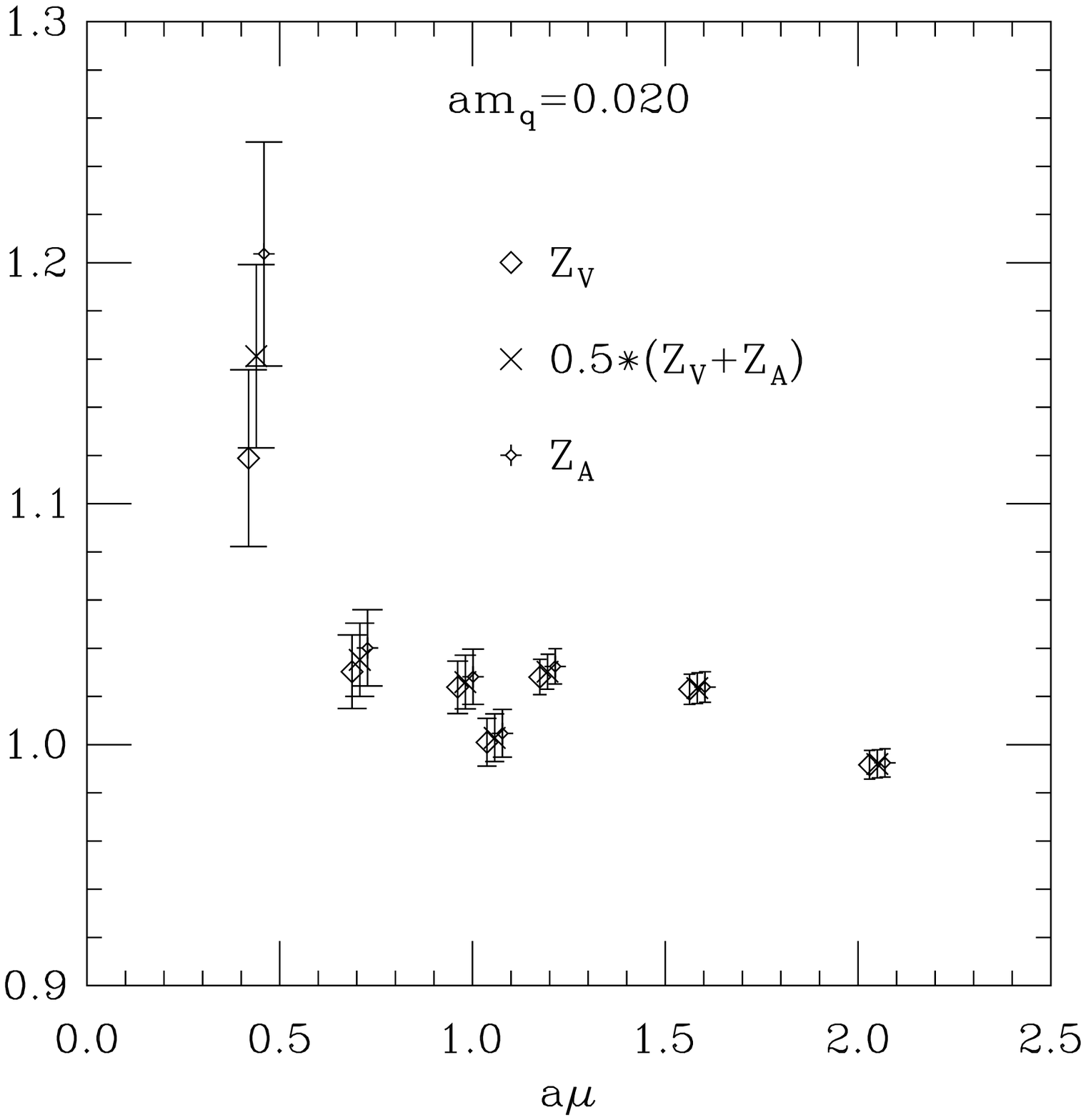}
\includegraphics[width=80mm]
{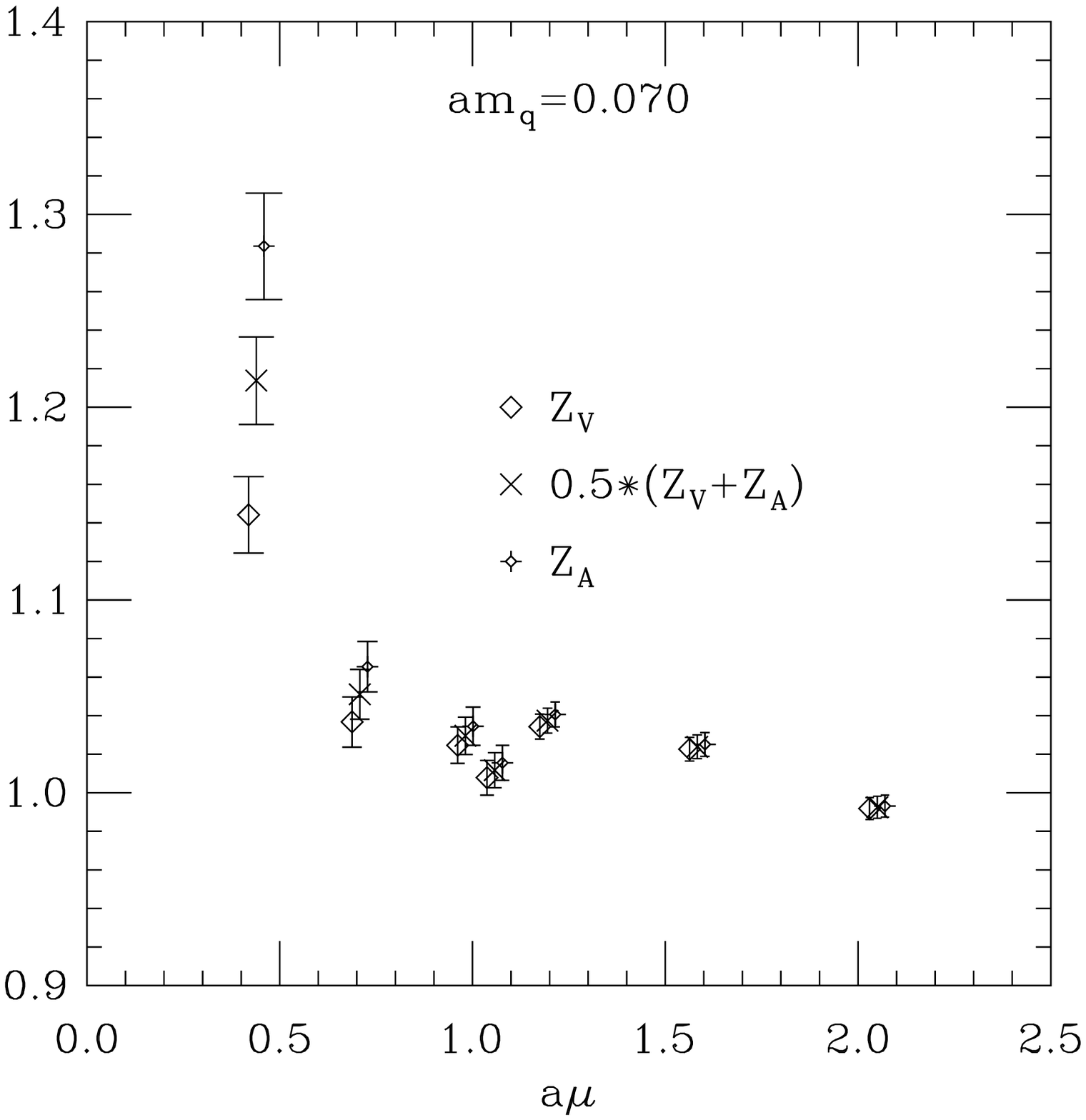}}
\caption{$Z_V$,$Z_A$ and their average versus $a\mu$ for quark masses
$am_q=0.020$ and 0.070. For clarity, at each value of the momentum $a\mu$,
the x-positions of $Z_V$ and $Z_A$ are shifted a little.}
\label{zvzapic1}
\end{figure}

\begin{figure}
{\centering\includegraphics[width=80mm]
{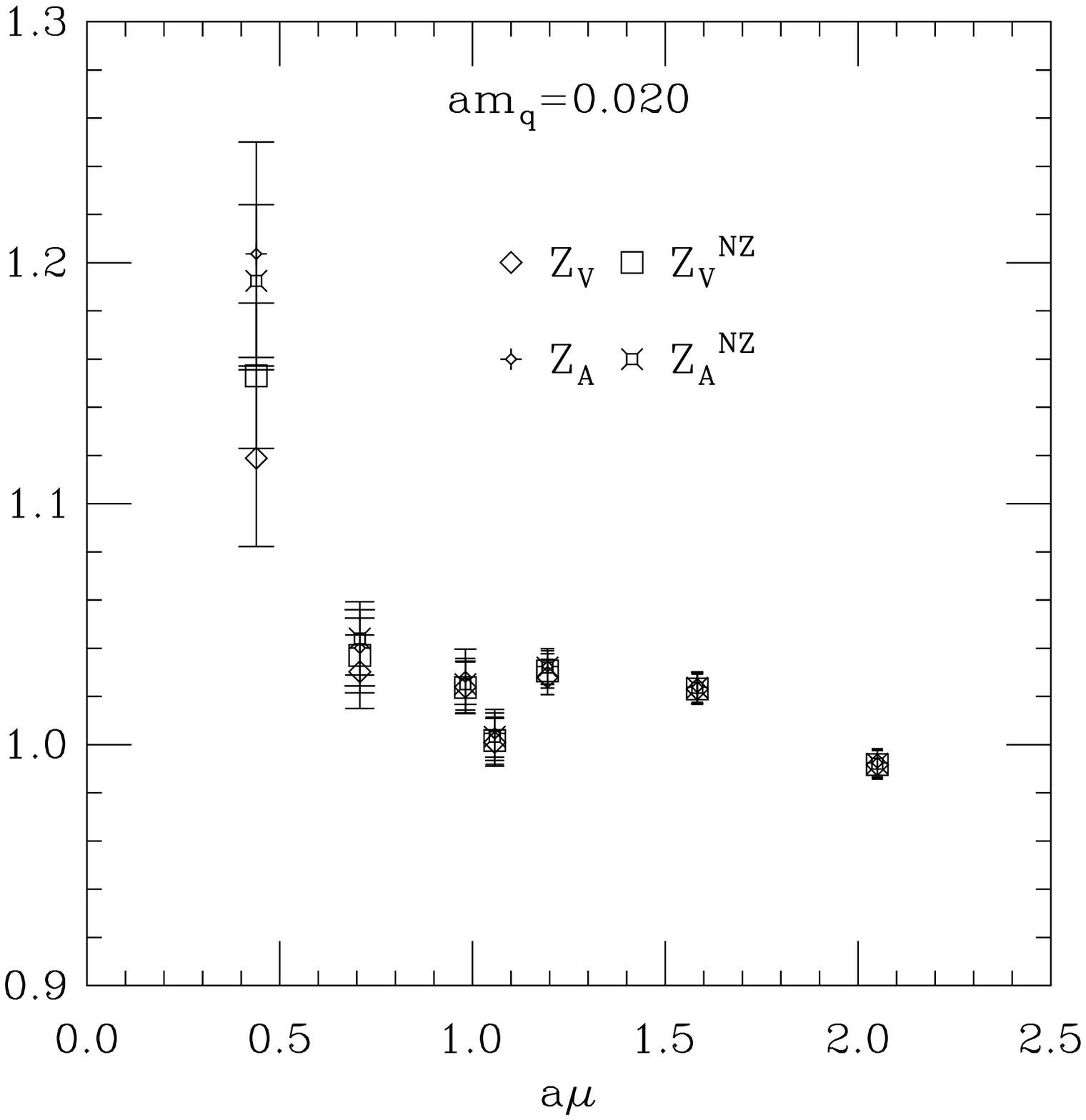}
\includegraphics[width=80mm]
{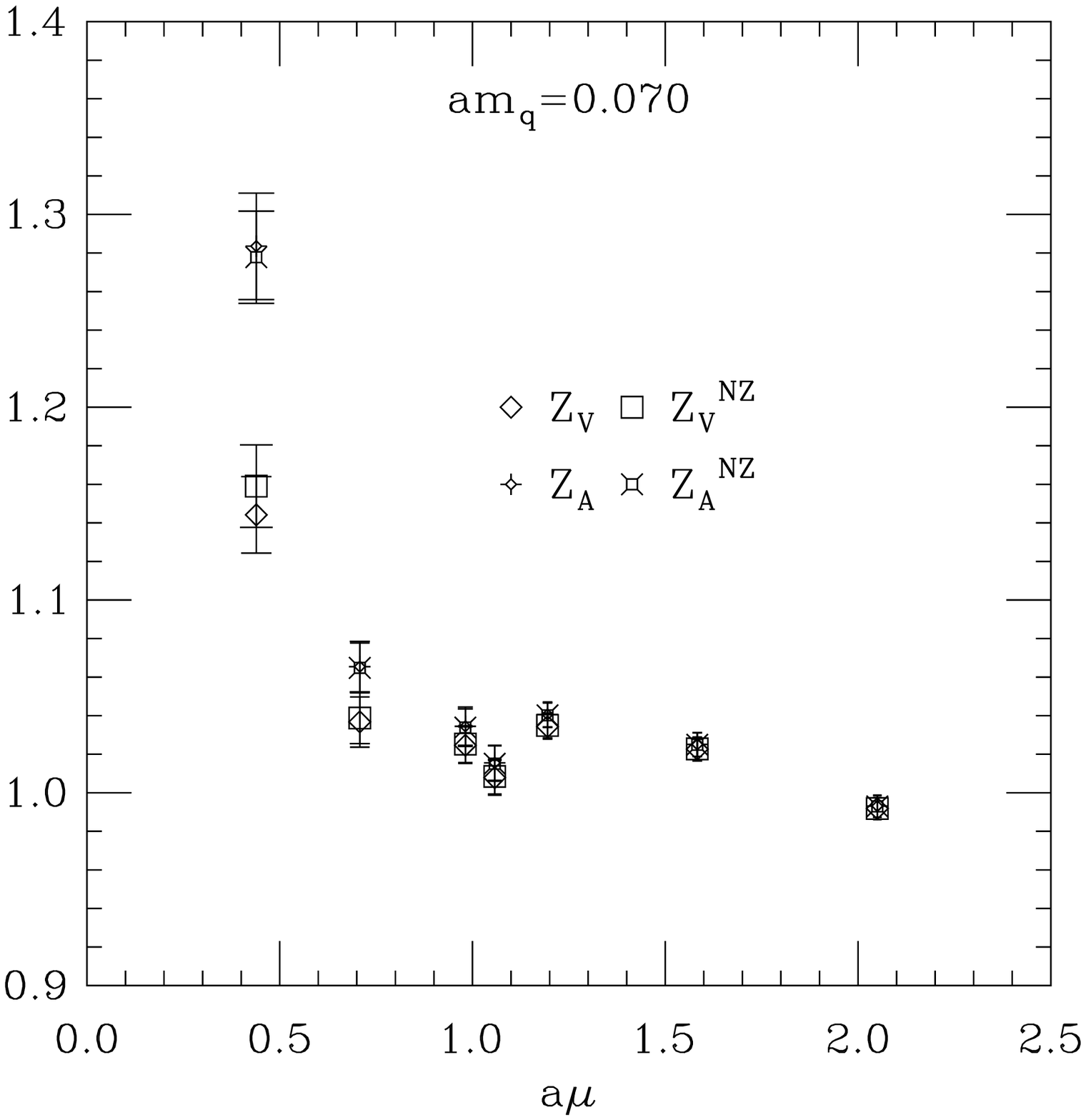}}
\caption{$Z_V^{NZ}$ and $Z_A^{NZ}$ compared with $Z_V$ and $Z_A$
for quark masses $am_q=0.020$ and 0.070.}
\label{zvza40n0}
\end{figure}

\begin{figure}
{\centering\includegraphics[width=80mm]
{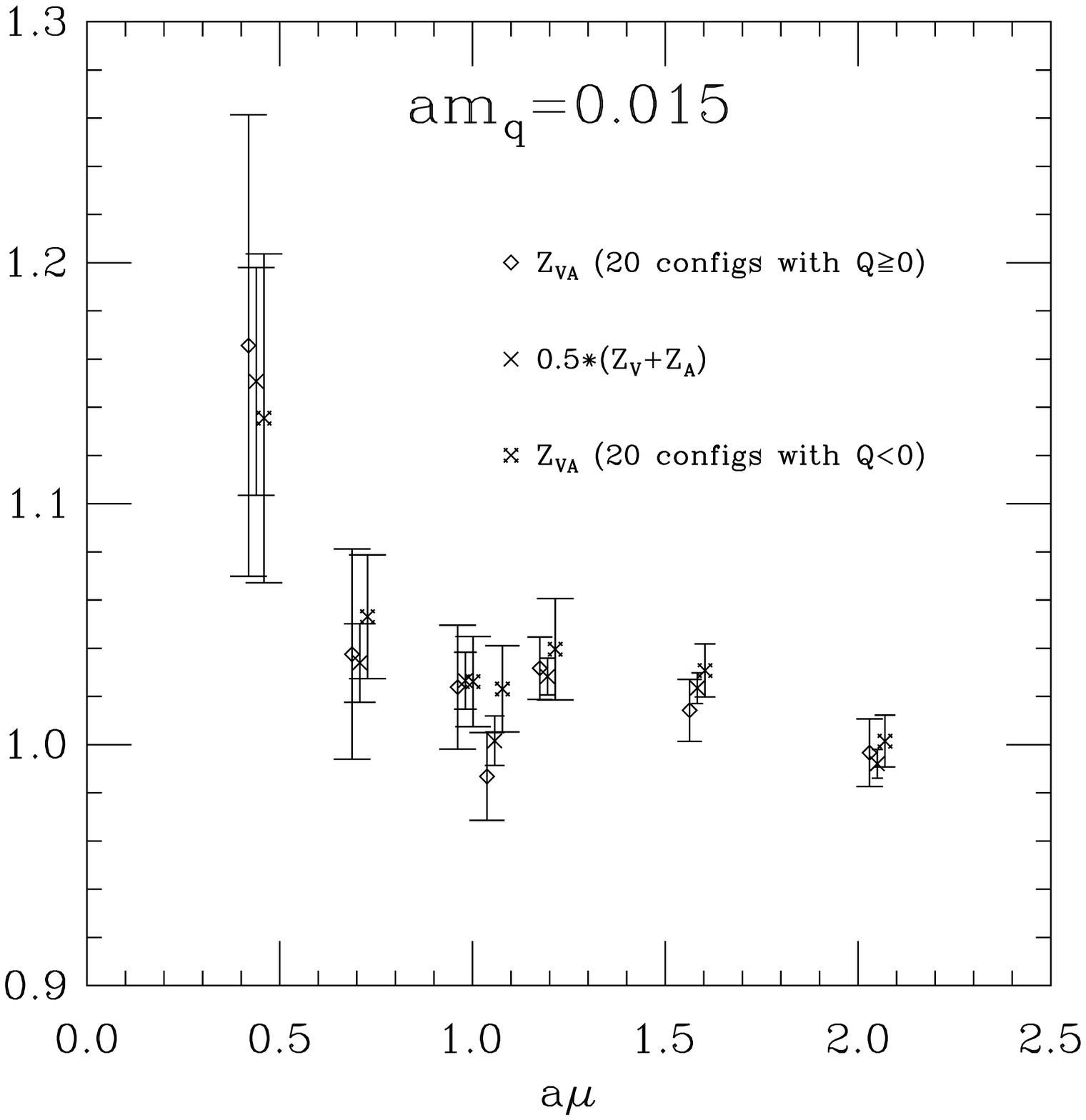}
\caption{$Z_{VA}$ and $0.5(Z_V+Z_A)$ at the smallest quark mass
$am_q=0.015$.}
\label{zvafig1}}
\end{figure}

In Fig.~\ref{zvzapic1}, the renormalization constants of the vector current
$Z_V$ and axial vector current $Z_A$ are given. For clarity,
at each value of the
momentum $a\mu$, the x-positions of $Z_V$ and $Z_A$ are shifted a little
in the graph.
As is shown in the graph, $Z_V$ and $Z_A$ are independent of the scale 
at large $\mu$ ($a\mu>0.7$), and $Z_V=Z_A$ within
statistical errors. At low $\mu$, $Z_A$ is bigger than $Z_V$. We think
it is because the axial vector current is coupled to the Goldstone
boson.
$Z_V^{NZ}$ and $Z_A^{NZ}$ obtained from quark
propagators with zero modes subtracted are compared with $Z_V$ and $Z_A$
in Fig.~\ref{zvza40n0}. Apparently, zero modes have little effect on
$Z_V$ and $Z_A$.
$Z_{VA}$ obtained from Eq.~(\ref{zva}) is shown in Fig.~\ref{zvafig1}.
It agrees with the average of $Z_V$ and $Z_A$. This confirms that
zero mode contribution doesn't matter in the computation of $Z_V$ and
$Z_A$.

\begin{figure}
{\centering\includegraphics[width=80mm]
{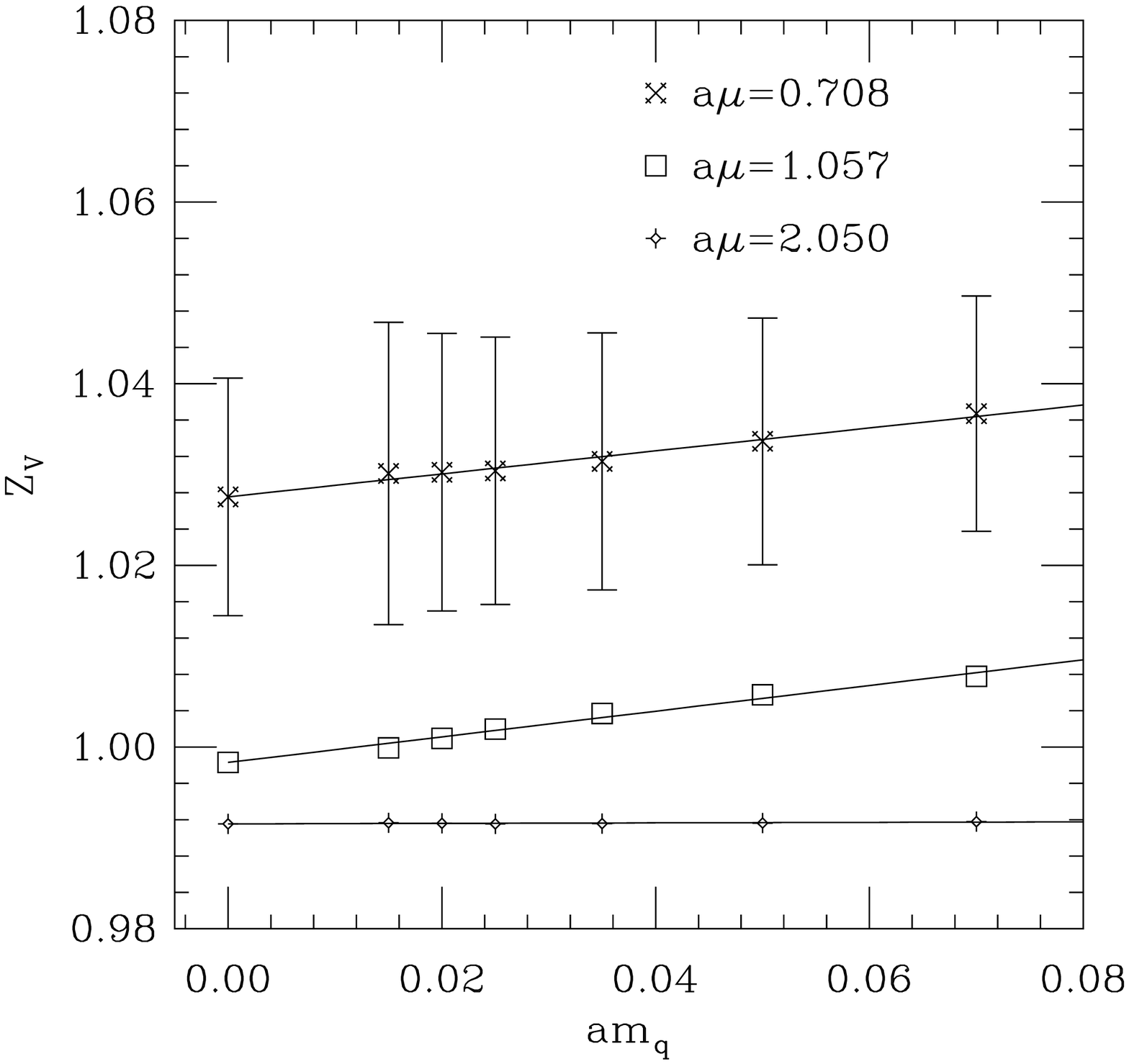}
\includegraphics[width=80mm]
{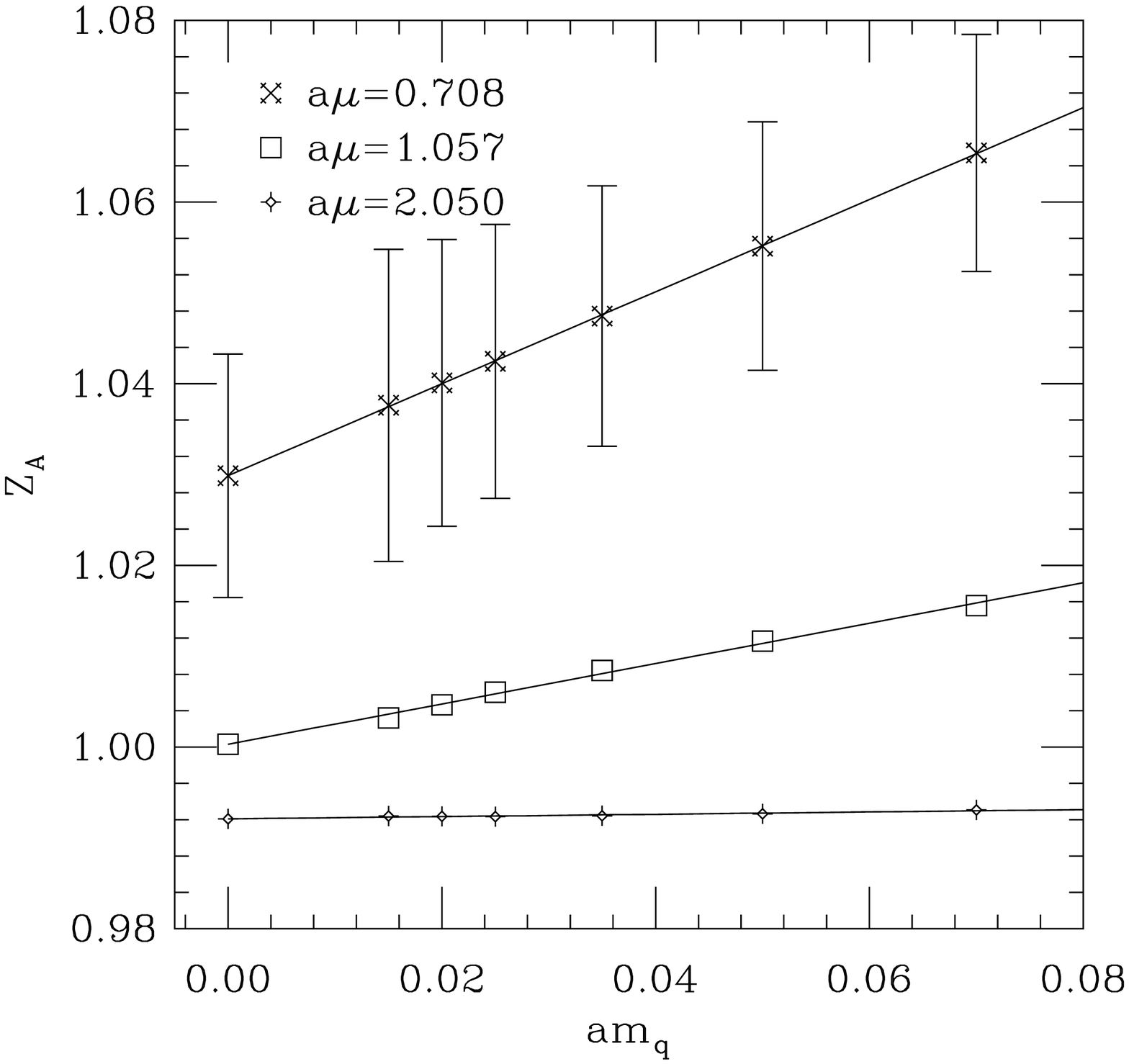}}
\caption{Extrapolation of $Z_V$ and $Z_A$ to the chiral limit.}
\label{zvandza}
\end{figure}

The RI' scheme values of $Z_V$, $Z_A$, $Z_V^{NZ}$ and $Z_A^{NZ}$
in the chiral limit are given in Table \ref{zvzatable}.
The superscripts $(a)$ and $(b)$ in the table indicate the two 
different ways of determining the lattice spacing.
The $\overline{MS}$ values are obtained by using the conversion
formulas in Section \ref{Conv}.
$Z_V=Z_A$ is very well satisfied as expected since the overlap
fermion respects chiral symmetry on the lattice.
The linear extrapolation to the chiral limit is shown in Fig.~\ref{zvandza}.
\begin{table}
\begin{center}
\begin{tabular}{|c|c|c|c|c|c|c|}
$a\mu$&$\mu^{(a)}$(GeV)&$\mu^{(b)}$(GeV)&$Z_V$&$Z_A$&$Z_V^{NZ}$&$Z_A^{NZ}$\\
0.439&1.08&0.96&1.11(3)&1.17(4)  &1.15(2) &1.16(3)\\
0.708&1.75&1.55& 1.03(1)&1.03(1)  &1.04(1) &1.04(1)\\
0.982&2.42&2.15&1.024(9)&1.026(10)&1.023(9)&1.022(9)\\
1.057&2.61&2.32&0.998(9)&1.000(9) &0.999(8)&0.999(8)\\
1.194&2.95&2.62&1.026(6)&1.029(6) &1.029(6)&1.029(6)\\
1.583&3.91&3.47&1.023(5)&1.023(6) &1.023(6)&1.023(6)\\
2.050&5.06&4.49&0.992(5)&0.992(5) &0.992(5)&0.992(5)\\
\hline\hline
$\overline{MS}^{(a)}$&2~GeV&&1.022(2)&1.023(1)&1.028(6)&1.028(7)\\
$\overline{MS}^{(b)}$&&2~GeV&1.021(1)&1.022(1)&1.022(4)&1.022(4)
\end{tabular}
\end{center}
\caption{Values of $Z_V$, $Z_A$, $Z_V^{NZ}$ and $Z_A^{NZ}$ 
in the RI' scheme
in the chiral limit. The $\overline{MS}$(2~GeV) value is obtained from
a linear interpolation from the two closest $\mu$ values of the data.
The lattice spacing is $(a)$:~0.08~fm from Sommer parameter or
$(b)$:~0.09~fm from the measured rho mass. Correspondingly, we get
two $\overline{MS}$(2~GeV) values.}
\label{zvzatable}
\end{table}

\section{Conversion to $\overline{MS}$}\label{Conv}
The ratio $Z_\Gamma^{\overline{MS}}(\mu^2)/Z_\Gamma^{RI'}(\mu^2)$ connects
the $\overline{MS}$ scheme to the RI' scheme is computed by continuum
perturbation theory. There is a need to determine the
coupling constant $\alpha_s(\mu)$ in the ratio. We obtain
$\alpha_s^{\overline{MS}}(\mu)$
by first measuring the trace of the plaquette operator $U_{plaq}$ (the
$1\times1$ Wilson loop), which can give us
$\alpha_s^V(3.41/a)$~\cite{tadpole}. Then $a\Lambda_V$ and 
$a\Lambda_{\overline{MS}}$ are calculated.
Finally $\alpha_s^{\overline{MS}}(\mu)$ is determined by
\begin{equation}
(\alpha_s^{\overline{MS}}(\mu))^{-1}=\beta_0\ln(\mu/\Lambda_{\overline{MS}})^2
+(\beta_1/\beta_0)\ln\ln(\mu/\Lambda_{\overline{MS}})^2,
\label{alphas}
\end{equation}
where $\beta_0=11/4\pi$ and $\beta_1=102/16\pi^2$ for the quenched
approximation. If the lattice spacing $a=0.08$~fm from the Sommer
parameter, we find
$\alpha_s^{\overline{MS}}(\mu=2$~GeV$)=\alpha_s^{\overline{MS}}(0.811/a)
=0.2038$. If $a=0.09$~fm from the measured rho mass, then
$\alpha_s^{\overline{MS}}(\mu=2$~GeV$)=\alpha_s^{\overline{MS}}(0.913/a)
=0.1940$.

For the scalar and pseudoscalar, in the Landau gauge and 3-loop order,
the conversion ratio is \cite{fran,chet}
\begin{eqnarray}
\frac{Z_S^{\overline{MS}}}{Z_S^{RI'}}=\frac{Z_P^{\overline{MS}}}{Z_P^{RI'}}
&=&1+\frac{16}{3}\frac{\alpha_s}{4\pi}+\left(\frac{4291}{18}
-\frac{152\zeta_3}{3}\right)\left(\frac{\alpha_s}{4\pi}\right)^2
\nonumber \\
&&+\left(\frac{3890527}{324}-\frac{224993\zeta_3}{54}
+\frac{2960\zeta_5}{9}\right)\left(\frac{\alpha_s}{4\pi}\right)^3
+O(\alpha_s^4),
\end{eqnarray}
where $\zeta_n$ is the Riemann zeta function evaluated at $n$.
Substituting $\alpha_s^{\overline{MS}}(\mu=2$~GeV$)=0.2038$ or 0.1940
into the above equation, we get
$Z_S^{\overline{MS}}/Z_S^{RI'}=Z_P^{\overline{MS}}/Z_P^{RI'}
=1+0.08650+0.04668+0.03131=1.1645$ or $1+0.08234+0.04230+0.02701=1.1516.$

For the vector and axial vector, since $Z_V^{RI}=Z_V^{\overline{MS}}$
and the difference between RI scheme and RI' scheme is only the
different definition of the quark field renormalization constants,
we have
\begin{equation}
\frac{Z_A^{\overline{MS}}}{Z_A^{RI'}}=\frac{Z_V^{\overline{MS}}}{Z_V^{RI'}}
=\frac{Z_V^{RI}}{Z_V^{RI'}}=\frac{Z_q^{RI'}}{Z_q^{RI}}
=\frac{Z_q^{RI'}/Z_q^{\overline{MS}}}{Z_q^{RI}/Z_q^{\overline{MS}}}
\end{equation}
$Z_q^{RI'}/Z_q^{\overline{MS}}$ and $Z_q^{RI}/Z_q^{\overline{MS}}$
were calculated in Ref.~\cite{fran,chet} to 3-loop, so we find
\begin{equation}
\frac{Z_A^{\overline{MS}}}{Z_A^{RI'}}=\frac{Z_V^{\overline{MS}}}{Z_V^{RI'}}
=1-\frac{67}{6}\left(\frac{\alpha_s}{4\pi}\right)^2-\left(
\frac{52321}{72}-\frac{607\zeta_3}{4}\right)
\left(\frac{\alpha_s}{4\pi}\right)^3+O(\alpha_s^4)
\end{equation}
The numerical value at $\mu=2$~GeV is $(1+0-0.00294-0.00232)=0.9947$
or $(1+0-0.00266-0.00200)=0.9953$.

In Table \ref{zszpsubtb1} and \ref{zvzatable}, 
the $\overline{MS}$ values at $\mu=2$~GeV
are obtained from linear interpolations between the two closest $\mu$
values of the data.

\section{Comparison with perturbative calculations}\label{Comp}
The perturbative calculation in Ref.~\cite{degr} gives the lattice to
$\overline{MS}$ matching factor
$Z_i=1+z_i\alpha_s(q^*)/3\pi$ at $a\mu=1$. Here i=S, P, V and A for
fermion bilinears. The values of
$z_i$ and the scale $q^*$ are given in Table V in Ref.~\cite{degr}.

We may use $\alpha_s^V(q^*)$ run from $\alpha_s^V(3.41/a)$, as is
determined from the plaquette~\cite{tadpole} or 
$\alpha_s^{\overline{MS}}(q^*)$ from Eq.~(\ref{alphas}). The results
of $Z_i$'s are listed in Table \ref{ptresults}. The ambiguity in the
choice of $\alpha_s$ and $q^*$ in perturbation
theory is small.
\begin{table}
\begin{center}
\begin{tabular}{c|cc|cc}
& $\alpha_s^V(1.96/a)$ & $\alpha_s^V(1.52/a)$ &
$\alpha_s^{\overline{MS}}(1.96/a)$ &
$\alpha_s^{\overline{MS}}(1.52/a)$\\
$Z_{S,P}$ & 1.010 & 1.011 & 1.008 & 1.009\\
\hline\hline
& $\alpha_s^V(1.26/a)$ & $\alpha_s^V(1.46/a)$ &
$\alpha_s^{\overline{MS}}(1.26/a)$ &
$\alpha_s^{\overline{MS}}(1.46/a)$\\
$Z_{V,A}$ & 0.989 & 0.990 & 0.991 & 0.992
\end{tabular}
\end{center}
\caption{Values of $Z_{S,P}$ and $Z_{V,A}$ at
$a\mu=1$ for HYP-planar overlap action 
from perturbative calculation in Ref.~\cite{degr}.}
\label{ptresults}
\end{table}
We have to run the result of $Z_{S,P}$
to $\mu=2$~GeV to compare with our $\overline{MS}$(2~GeV) value.
We use the two loop formula for the running quark mass given in
Ref.~\cite{bura}(Eq.~(4.81)).
If the inverse lattice spacing is $1/a=2.47$~GeV ($a=0.08$~fm) from
the Sommer parameter, then we 
find $Z_{S,P}(2$~GeV$)=0.975$ from
$Z_{S,P}(2.47$~GeV$)=1.009$.
If the inverse lattice spacing is $1/a=2.19$~GeV ($a=0.09$~fm) from 
the measured rho mass, then $Z_{S,P}(2$~GeV$)=0.995$.
In any case, the value of $Z_{S,P}$
from perturbative calculation is quite close to 1, while our non-perturbative
results 0.79(5)/0.89(4) or 0.83(4)/0.93(3) (see Table \ref{zszpsubtb1}) 
are not.
Thus, perturbative calculation of the matching factors for scalar
and pseudoscalar density for HYP-planar overlap action seems unreliable.

Unlike $Z_S$ or $Z_P$, $Z_V$ and $Z_A$ are scale independent.
We can compare the values of $Z_{V,A}$ in Table
\ref{ptresults} directly with our non-perturbative $\overline{MS}$(2~GeV)
results in Table \ref{zvzatable}. All are quite close to one (the shift
from one is less than 0.03). This indicates that we can believe in
the perturbative calculations of $Z_V$ and $Z_A$ for the HYP-planar overlap
action.

\section{Summary and conclusion}\label{Summ}
We calculated the renormalization constants of bilinear quark operators
non-perturbatively using the HYP-planar overlap action with exact
chiral symmetry. By comparing the results with those from perturbative
computation, we find that a perturbative calculation is reliable with $Z_V$ and
$Z_A$, but not with $Z_S$ and $Z_P$. The exact zero modes of the Dirac
operator turn out to be important in calculating $Z_S$ and $Z_P$, while
not relevant in calculating $Z_V$ and $Z_A$. After subtracting
the zero modes from the quark propagator, $Z_S=Z_P$ is well satisfied.
$Z_V$ and $Z_A$ are also in good agreement with each other as is
expected from the chiral symmetry of the action.
We expect that zero modes will be much less important in simulations done with
dynamical overlap quarks\cite{DeGrand:2004nq}.

The perturbative result that actions using HYP-blocked links have
matching factors quite close to unity is confirmed for vector and
axial vector currents with our HYP-planar overlap action. This does not
appear to be the case for the scalar and pseudoscalar densities.

\section*{Acknowledgments}
                                                                                
This work was supported by the US Department of Energy.

\end{document}